\documentclass[aps,pre,preprint,groupedaddress,showpacs]{revtex4-2}
\usepackage{graphicx,amsmath,amssymb,xcolor}
\usepackage[normalem]{ulem}
\fontfamily{qhv}\selectfont
\usepackage{capt-of}
\usepackage[font=sf]{caption}
\usepackage{setspace}
\newcommand{\z}[1]{{\color{black}#1}}

\begin{document}

\title{Time evolution of the Boltzmann entropy for a nonequilibrium dilute gas}
\author{Pedro L. Garrido}
\affiliation{Institute Carlos I for Theoretical and Computational Physics, Universidad de Granada, Spain}
\author{Sheldon Goldstein}
\affiliation{Department of Mathematics, Rutgers University, Piscataway, NJ, 08854, USA}
\author{David A. Huse}
\affiliation{Department of Physics, Princeton University, Princeton, NJ, 08544, USA}
\author{Joel L. Lebowitz}
\affiliation{Departments of Mathematics and Physics, Rutgers University, Piscataway, NJ, 08854, USA}

\begin{abstract}
We investigate the time evolution of the Boltzmann entropy of a dilute gas of $N$ particles, $N\gg 1$, as it undergoes a free expansion doubling its volume. The microstate of the system 
changes in time via Hamiltonian dynamics.  
Its entropy, at any time $t$, is given by the logarithm of the phase space volume of all the microstates giving rise to its macrostate at time $t$. The macrostates that we consider are defined by coarse graining the  
one-particle phase space into cells $\Delta_\alpha$. The initial and final macrostates of the system are thermal equilibrium states in volumes $V$ and $2V$, with the same energy $E$ and particle number $N$. Their entropy per particle is given, for sufficiently large systems, by the thermodynamic entropy as a function of the particle and energy density, whose leading term is independent of the size of the $\Delta_\alpha$. The intermediate (non-equilibrium) entropy does however depend on the size of the cells $\Delta_\alpha$.  Its change with time is due to (i) dispersal in physical space from free motion and to (ii) the collisions between particles which change their velocities.  The former depends strongly on the size of the velocity coarse graining $\Delta v$: it produces entropy at a rate proportional to $\Delta v$. This dependence is investigated numerically and analytically for a dilute two-dimensional gas of hard discs.  It becomes significant when the mean free path between collisions is of the same order or larger than the length scale of the initial spatial inhomogeneity.  In the opposite limit, the rate of entropy production is essentially independent of $\Delta v$ and is given by the Boltzmann equation for the limit $\Delta v\rightarrow 0$. We show that when both processes are active the time dependence of the entropy has a scaling form involving the ratio of the rates of its production by the two processes.
\end{abstract}
\maketitle

\section{ Introduction} 
Consider an isolated macroscopic system of $N$ particles, $N\gg 1$, in equilibrium in a region $\Lambda$ of $d$-dimensional space with volume $|\Lambda|=V$.  Removing a constraint at time $t=0$ permits the system (still isolated) to expand into a larger region $\Lambda'$, with volume $|\Lambda'|=V' > V$.  After some sufficiently long time the system will come to a new equilibrium in $\Lambda'$, with the same total energy and particle number.  The second law of thermodynamics states that the entropy of the new equilibrium system is greater than what it was originally, i.e.
\begin{equation}
S_{eq}(E, N,  V') > S_{eq}(E, N, V) ~ , ~V' > V~.
\end{equation}
The thermodynamic equilibrium entropy for macroscopic systems, $S_{eq}(E, N, V)$, is an extensive well-defined quantity (up to an additive constant) first introduced by Clausius in 1857 \cite{clausius}. 


Clausius does not say anything explicit about the entropy of the system while it is in a nonequilibrium state, as in the above case while it is transitioning from its initial to its final equilibrium state.  Boltzmann, looking for 
a justification of the second law, which seems to contradict the reversibility of the microscopic dynamics, came to the brilliant insight: 
the entropy of a macroscopic system in a microstate $X= ({\bf r}_1, {\bf v}_1, \cdots , {\bf r}_N, {\bf v}_N)$, giving rise to macrostate $M$, \z{corresponding to the values of suitable macrovariables,} 
is proportional to $\log|\Gamma_M|$.   
$\Gamma_M$ is the region of the $(2dN)$-dimensional phase space of the system all of whose microstates $X$ \z{are macroscopically similar in the sense that they }give rise to this same macrostate $M$, and $|\Gamma_M|$ is its Liouville volume.  This applies for equilibrium and nonequilibrium micro- and macrostates.  
We shall denote this Boltzmann entropy by $S_B (X) \equiv S_B (M (X))=\log{|\Gamma_{M(X)}|}$ \cite{pen,leb}; see also \cite{obsent}.  As the microstate $X$ evolves in time from a nonequilibrium macrostate $M(X)$, $|\Gamma_{M(X(t))}|$ typically increases.  Thus this $S_B(X(t))$ satisfies the second law for the vast majority of microstates $X$ in each macrostate $M$ without violating microscopic reversibility. [Note, in this paper we are using units where Boltzmann's constant is $k_B = 1$.]

The choice of macrostates $M$, which corresponds to dividing each energy shell into regions $\Gamma_M$, is not unique but is physically constrained. In particular we want there to be an equilibrium macrostate $M_{eq}$ such that $S_B(M_{eq})\cong S_{eq}$ when the size of the system goes to infinity.  A common way to choose $M$ is to divide the spatial region $\Lambda  \in \Bbb R^d$ into cells that are each large enough to contain many particles and specify ``within some tolerance'' the total energy,  total particle number, and total momentum in each such cell.  These are the locally conserved quantities corresponding to the hydrodynamical variables which, in most cases, evolve on a macroscopic time scale according to autonomous equations, e.g. the Navier-Stokes equations \cite{spohn}.  


Going beyond hydrodynamical variables, Boltzmann also considered, 
for dilute gases, 
more refined macrostates, $M_g$, than those given by the spatial profiles of the hydrodynamic variables.  
He defined $M_g$ by considering the six-dimensional {\it single-particle} phase space, $\gamma = \{ \mathbf{r}, \mathbf{v}\},~ \mathbf{r} \in \Lambda \in \Bbb R^3, ~\mathbf{v} \in \Bbb R^3$.
The microstate $X$ of the system is described by $N$ points in $\gamma$, while the macrostate $M_g(X)$ is specified by dividing $\gamma$ into cells $\Delta_\alpha$ and giving, within some tolerance, the number of particles $N_\alpha$ in each cell $\alpha$.  The $N_\alpha$ satisfy the conditions:  
\begin{equation}
\sum_\alpha N_\alpha = N, \z{{~{\sum_{\alpha}\frac12 N_{\alpha}|{\bf v_{\alpha}}|^2\cong E,\ }}}
\end{equation}
with ${\bf v}_{\alpha}$ the mean velocity 
in the cell $\Delta_{\alpha}$; the particles each have mass $m=1$.  The potential energy is assumed to be negligible, 
although the particle-particle scattering due to the interactions is not neglected.  This is appropriate only for dilute gases.  For more general systems we also have to specify the potential energy, {\it c.f.} \cite{gl}.

Boltzmann then computed $|\Gamma_{M_g}|$ to be proportional to
\begin{equation}
|\Gamma_{M_g}|\sim {\bf \Pi}_\alpha [\frac{|\Delta_\alpha|^{N_\alpha}}{ N_\alpha!}] ~,
\label{eq:GammaM}
\end{equation}
from which he obtained the entropy $S_B(M_g(X))=\log{|\Gamma_{M_g}|}$.  To obtain a truly macroscopic description of the system, the cells should be large enough so that most particles are in cells with $N_{\alpha}\gg 1$.   Using Stirling's formula, Boltzmann then obtained
\begin{equation}
S_B(M_g)=\log{|\Gamma_{M_g}|}  = -\sum_\alpha |\Delta_\alpha| [\frac{N_\alpha}{|\Delta_\alpha|}\log{\frac{N_\alpha}{|\Delta_\alpha|}}]+{\rm constant}~,
\label{eq:SB}
\end{equation}
where the constant depends on $N$.  When we give formulae for entropies later in this
paper, we leave off this additive constant, giving only the part of the entropy that depends on the
configuration of the particles.

\section{The One-Dimensional Ideal Gas}

In a previous work with other collaborators, two of us investigated the time evolution of the $\{N_\alpha (t)\}$, and thus of $S_B(M_g(X(t)))$, as given in (\ref{eq:SB}), for an ideal gas in one dimension \cite{dhar} (see also De Bievre and Parris \cite{Bievre}). We chose cells $\Delta_{\alpha}$ all of equal size $(\Delta x\Delta v)$ (with a cutoff on the maximal speed $|v|$).  We started the system in a thermal equilibrium microstate, confined in an interval of length $L$, and then let it freely expand to fill an interval of length $2L$.  After the system equilibrates, {\bf $X(t)\in\Gamma_{M_{eq}}$ in the larger interval at almost all times, and} the entropy $S_B(M_g) \simeq S_{eq}(2L)$, essentially independent of the choices of cell sizes $\Delta x$, $\Delta v$.  During this process the entropy has increased by (approximately) $\log{2}$ per particle due to the expansion.  We say approximately because we are not in the limit $N\rightarrow\infty$, $L\rightarrow\infty$.  [This change, $\log{2}$, is for classical dilute gases.  For quantum gases, on the other hand, the change depends on the initial temperature and on the particle statistics (fermions {\it vs.} bosons)  
\cite{dgm,kardar,quantum}.]

We found that  the ``equilibration'' time, $t_{eq}$, it took the Boltzmann entropy of the system to approach the new equilibrium value in the final interval of length $2L$ depended strongly on the width $\Delta v$ of the single-particle phase space cells used to define the macrostate $M_g$, see Fig. {\ref{figi1}}. 
The smaller $\Delta v$, the slower the rate of entropy production for this $S_B(M_g)$. 
{\bf As a consequence of this, the Boltzmann entropy of each particular nonequilibrium microstate $X(t)$ that occurs during this free expansion is a strongly varying function of the chosen $\Delta v$.}

In fact we found and proved that $t_{eq} \sim L/\Delta v$ for small $\Delta v$.  The reason for this is that in the ideal gas the only mechanism for uniformising the velocity distribution over all of the spatial region is via the difference between the total distance traveled in time $t$ by the particles with velocity $v$ and those with velocity $(v+\Delta v)$.  In order for the system to approach equilibrium (become spatially uniform), this distance must exceed $L$, which only occurs after time $t_{eq}\sim L/\Delta v$.  Note that this equilibration time $t_{eq}$ diverges in the limit $\Delta v\rightarrow 0$.  

The time evolution of the entropy of the ideal gas is much less sensitive to the spatial size $\Delta x$ of the cells. 
The time scale for the uniformization of the spatial density (ignoring the local velocity distributions) is of order $L/v_{th}$, where $v_{th}$ is the (thermal) mean speed, so for $\Delta v\ll v_{th}$ this time is much smaller than $t_{eq}$, see Fig. {\ref{fig1}}.

These observations are consistent with, \z{and in fact imply, 
the non-increase of entropy} in the limiting case 
in which $\Delta x\rightarrow 0$, $\Delta v\rightarrow 0$, and $N\rightarrow\infty$, such that most particles are in boxes with $N_{\alpha}\gg 1$, while 
$\{N_\alpha(t)/ (N|\Delta_\alpha|)\}\to f(x, v,t)$, a piecewise smooth function.  
\z{To directly see this non-increase, note that} since we have an ideal gas, the time evolution of this smooth empirical distribution satisfies the equation 
\begin{equation}
\frac{\partial f (x, v, t)}{\partial t} + v \frac{\partial f(x,v,t)}{\partial x} = 0~. 
\label{eq:free}
\end{equation}
Taking the corresponding limit of the entropy per particle $S_B(M_g)/N$ given in Eq. ({\ref{eq:SB}}) yields \cite{spohn} 
\begin{equation}
s(f_t) = - \int dx \int dv~f (x, v, t) \log f (x, v, t) ~,
\label{eq:sbf} 
\end{equation}
where we do not show a time-independent additive constant.  As is well known, $s(f_t)$ is time-invariant under the ideal gas evolution given by (\ref{eq:free}); the entropy production due to $\Delta v>0$ and free particle motion vanishes when we take the limit $\Delta v\rightarrow 0$.  This shows that $s(f_t)$, as defined by (\ref{eq:sbf}), corresponding to the volumes $|\Delta_{\alpha}|\rightarrow 0$, does not change with time for an ideal gas. 

\z{Note, however, that when $f(x,v,t)$ describes, not the {\it empirical} distribution of the phase point of our gas in the limit described above, but rather the one-particle distribution  of an {\it ensemble} of fixed finite systems of independent particles, the non-increase of the corresponding ensemble entropy per particle $s(f_t)$ conflicts dramatically with the increase of the Boltzmann entropy $S_B(M_g(X(t)))$ of the system. We stress that, unlike the former, the latter depends on the choice of $\Delta v$, with  the rate of change of $S_B$ decreasing as $\Delta v$ is decreased. }

\section{The Dilute Gas}

The question then arises of what happens to the Boltzmann entropy $S_B(M_g)$ for different cell sizes $|\Delta_\alpha|$ when one takes into account interactions between the particles.  With such interactions there is a mechanism  
for changing the velocity of a particle, so Eq. (\ref{eq:free}) and its generalization to more than one dimension no longer describe the time evolution of the limiting single-particle empirical distribution in $\{{\bf r, v}\}$ space.  Let us consider in particular the case of a dilute gas in two or three dimensions with short-range interactions, such as hard discs or spheres.  (See also Ref. \cite{cdk} for a related study of a two-component interacting gas in one dimension.)  
For a dilute gas in two or more dimensions, 
Lanford proved that in the Boltzmann-Grad (B-G) limit $f({\bf r, v},t)$ evolves (for short times) according to the Boltzmann equation (BE) \cite{lanford,bgs, spohn} 
\begin{equation}
\frac{\partial f({\bf r,v},t)}{\partial t} + {\bf v\cdot\nabla} f({\bf r,v},t) = \lambda^{-1} Q (f, f) ~;  
\label{eq:bolt}
\end{equation}
the right hand side of this equation describes the interparticle collisions.  In this B-G limit the particle density goes to infinity while the diameter of the hard spheres vanishes in such a way that the mean free path (mfp) between collisions, $\lambda$, remains fixed.  $f({\bf r}, {\bf v}, t)$ in (7) is
then exactly the smooth density profile of the empirical distribution when $|\Delta_\alpha|\rightarrow 0$.

The Boltzmann equation (\ref{eq:bolt}) derived by Boltzmann on the basis of physical arguments is known to describe the ``smoothed'' empirical single-particle distribution $f({\bf r,v},t)$ of a dilute gas for the case where the size of the atoms is very small compared to the interparticle distance which in turn is very small compared to $\lambda$ \cite{cercignani,cip}. 

Boltzmann proved that $s(f_t)$ defined in (\ref{eq:sbf}) is monotone increasing with time $t$ when $f({\bf r,v},t)$, given by the solution of (\ref{eq:bolt}), is not equal to the local Maxwell-Boltzmann distribution, his famous $H$-theorem. 
Boltzmann saw this as a generalization of the second law. He wrote that with the $H$-theorem ``we are able to generalize the notion of entropy to nonequilibrium systems''  [p. 75 in \cite{bolt}].

Boltzmann did not seem to worry about the fact that  
$s(f_t)$ does not increase with time for the ideal gas, when $\lambda\to\infty$. In fact, 
even when $\lambda$ is finite  
there is a nonzero contribution from the free-particle dispersion to the rate of increase with time of  
$S_B(M_g)$, for $|\Delta_{\alpha}|>0$. This comes from  
the term ${\bf v \cdot\nabla} f$ in the BE. 
This contribution is small when $\lambda$ is small compared to the length scale of the spatial inhomogeneity {\bf as was certainly the case for the gases Boltzmann considered}, it is zero for a spatially uniform system, but it can be substantial 
when $\lambda$ is of the order of or larger than the linear scale of the inhomogeneities.  This scale is of order the linear size $L$ of the system in the case of the free expansion considered here.  The dependence of $S_B(M_g(t))$ on $\lambda$ and on the choice of cell sizes is the question we address next for a dilute interacting gas.

We note that Eq. (\ref{eq:free}) for the limiting empirical distribution, $f(x,v,t)$, is obtained formally for $|\Delta_\alpha|\rightarrow 0$ by following Lanford's steps in the derivation of the BE, if we take a limit where the diameter goes to zero faster than in the B-G limit so that the mean free path goes to infinity.  

\section{Hard discs}

To elucidate the time evolution of $S_B(M_g(t))$ for different choices of $|\Delta_{\alpha}|$ when $\lambda$ is of the same order or larger than the length scale of the inhomogeneities, we have carried out molecular dynamics computations for the time evolution of a two-dimensional system of $N$ hard discs of unit mass.  The system is started in a microstate chosen at random from a canonical Gibbs ensemble with temperature $T=1$ (setting $k_B=1$) in a rectangular box of size $L_x=1/2$, $L_y=1$ with periodic boundary conditions along the $y$ direction and hard walls constraining the system along the $x$ direction. The discs have radius $r$. 
The system has initial areal density $\eta^{(0)}=\pi r^2 N/V$ where $V=L_xL_y=1/2$.  The corresponding mean free path when the system is dilute is $\lambda\sim (\eta N/V)^{-1/2}$.

At time $t=0$ we remove the hard walls and let this gas of discs expand to a box of size $L_x=L_y=L=1$ with now periodic boundary conditions along both directions (see figure \ref{fig1}). We study the time evolution of this 
system until it reaches an equilibrium state $X\in \Gamma_{M_{eq}}$ in this larger periodic box, with final areal density $\eta=\eta^{(0)}/2$.  Although in the simulations we have chosen to use $L=1$, in much of the analysis below we will show the dependence on $L$, for generality.  

\begin{center}
\includegraphics[width=5cm]{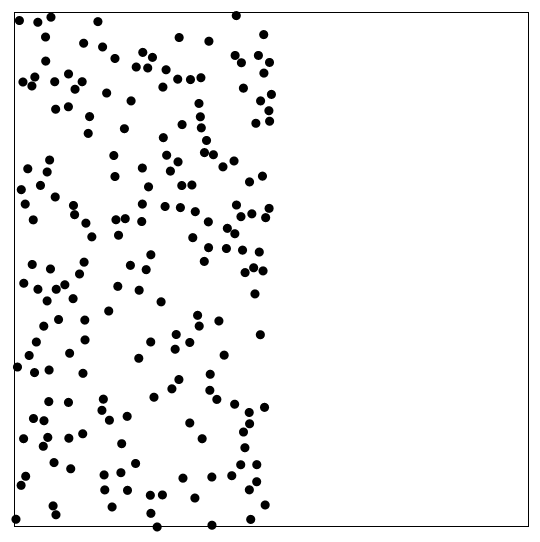}  
\includegraphics[width=5cm]{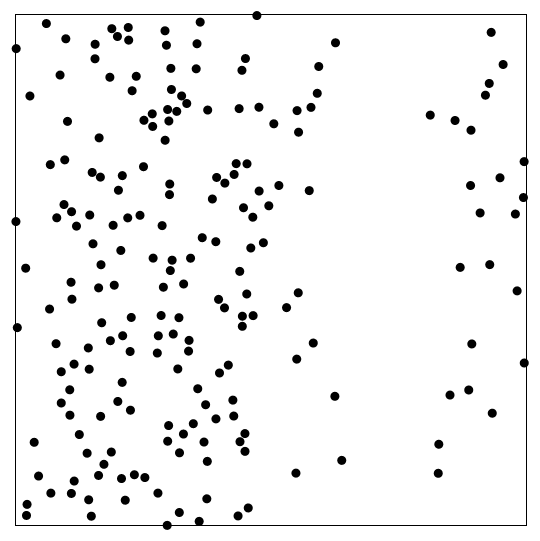}
\includegraphics[width=5cm]{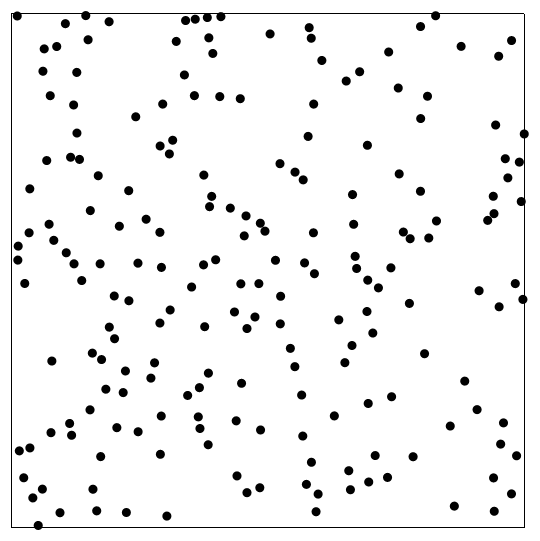}
\captionof{figure}{Evolution of a system with $N=200$ hard discs, initial areal density $\eta^{(0)}=0.1$ and initial temperature $T=1$: $t=0$, $0.1$,  and $0.3$ from left to right.} 
\label{fig1}  
\end{center}

In the regime $\lambda\gtrsim L$ the degrees of freedom associated with $y$ and $v_y$, along which direction the system does not expand, remain near thermal equilibrium.  Therefore the Boltzmann entropy associated with these $y$, $v_y$ degrees of freedom remains approximately constant in time while the system expands along the $x$ direction and approaches the new thermal equilibrium.  Thus we will focus only on the entropy due to the degrees of freedom associated with $x$ and $v_x$, since it is only this part of the entropy that is strongly out of equilibrium and changing with time.  For the ideal gas, $\lambda\rightarrow\infty$, this separation of $y$ degrees of freedom staying at thermal equilibrium, while the $x$ degrees of freedom do not, becomes exact.

Thus we divide the four-dimensional one-particle phase space $(x,v_x;y,v_y)$ into cells $\Delta_{\alpha}$ that are all of extent $\Delta x$ and $\Delta v_x$ along the $x$ and $v_x$ directions, respectively, with each cell including the full range of $y$ and $v_y$.  
We count $N_{\alpha}(t)$, the number of particles in $\Delta_\alpha$ at time $t$.  Following Boltzmann in using Stirling's formula for $\log{|\Gamma_M|}$ in (\ref{eq:SB}), the Boltzmann entropy per particle due to their $x$ and $v_x$ degrees of freedom is:
\begin{equation}
\frac{1}{N}S_B(M_g(t))=s_B(t)= -\frac{\Delta}{N}\sum_{\alpha}  \frac{N_{\alpha}(t)}{\Delta}\log \frac{N_{\alpha}(t)}{\Delta}~. 
\label{eq:entz}
\end{equation}
In (\ref{eq:entz}) all cells have equal ``area'' $\vert\Delta_{\alpha}\vert=\Delta x\Delta v_x=\Delta$ in the $(x,v_x)$ phase space. 


To specify $\Delta_\alpha$ 
we divide $x \in [0,L]$ into $n_x$ equal intervals. We used different values of $n_x$: $n_x=4,8,16$; as in \cite{dhar} the results show little dependence on $\Delta x$, so here we show only the results for $n_x=16$.  Similarly, we divide the range of the velocity $v_{x}\in [-v_{max},v_{max}]$, with $v_{max}=6\sqrt{T}=6$, into $n_v=4,8,\ldots,256$ equal cells, so $\Delta v_x=12/n_v$. 
Again, our cells divide the single-particle phase space only along $x$ and $v_x$, so each cell includes the full ranges of $y$ and $v_y$.

\subsection{Ideal Gas; $\lambda=\infty$}
Before describing the results for different finite values of $\lambda$, we present the time evolution for the case of $\lambda=\infty$, i.e., the ideal gas, in Fig. {\ref{figi1}}. These figures closely resemble the one-dimensional ideal gas case studied in \cite{dhar}. We observe how the equilibration time of the Boltzmann entropy increases as $\Delta v_x$ decreases. Once we scale time as $\tau=\Delta v_x t/L$, ($L=1$), we observe the convergence of the curves for different $\Delta v_x$ values towards a limiting curve as $\Delta v_x\rightarrow 0$. The limiting curve for the entropy was obtained in reference \cite{dhar} from the ideal gas equation (\ref{eq:free}). 
The deviations from this limiting behavior become substantial when $\Delta v_x$ is of order $v_{th}$, which is the case for our largest $\Delta v_x$.

Note that there is a slight, rapid increase in entropy at very early times, attributed to the initial stages of the expansion, see Fig. {\ref{figi1}}. The specific behavior in this early phase is influenced by the choice of $\Delta x$ and the arrangement of cell boundaries along $x$, with two of them precisely at the locations of the initial confining walls. However, our focus will be on the behavior at intermediate and late times, which are not influenced by these choices associated with the cells along the $x$ direction.
\begin{center}
\includegraphics[height=7cm]{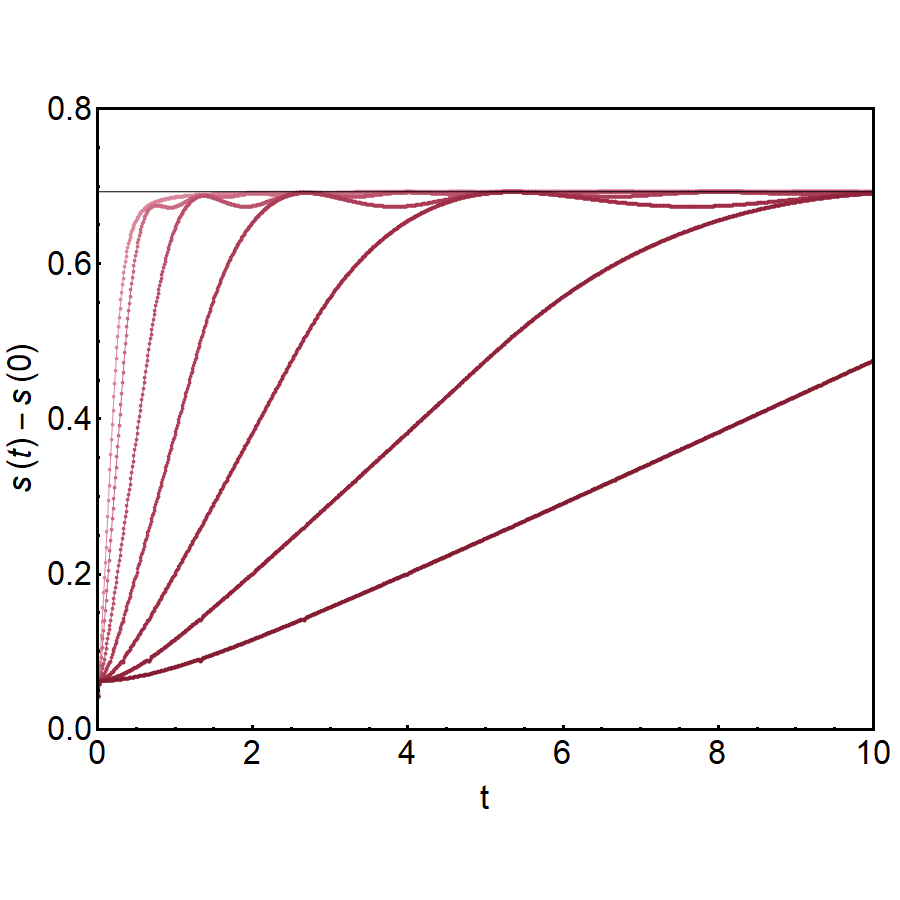}  
\includegraphics[height=7cm]{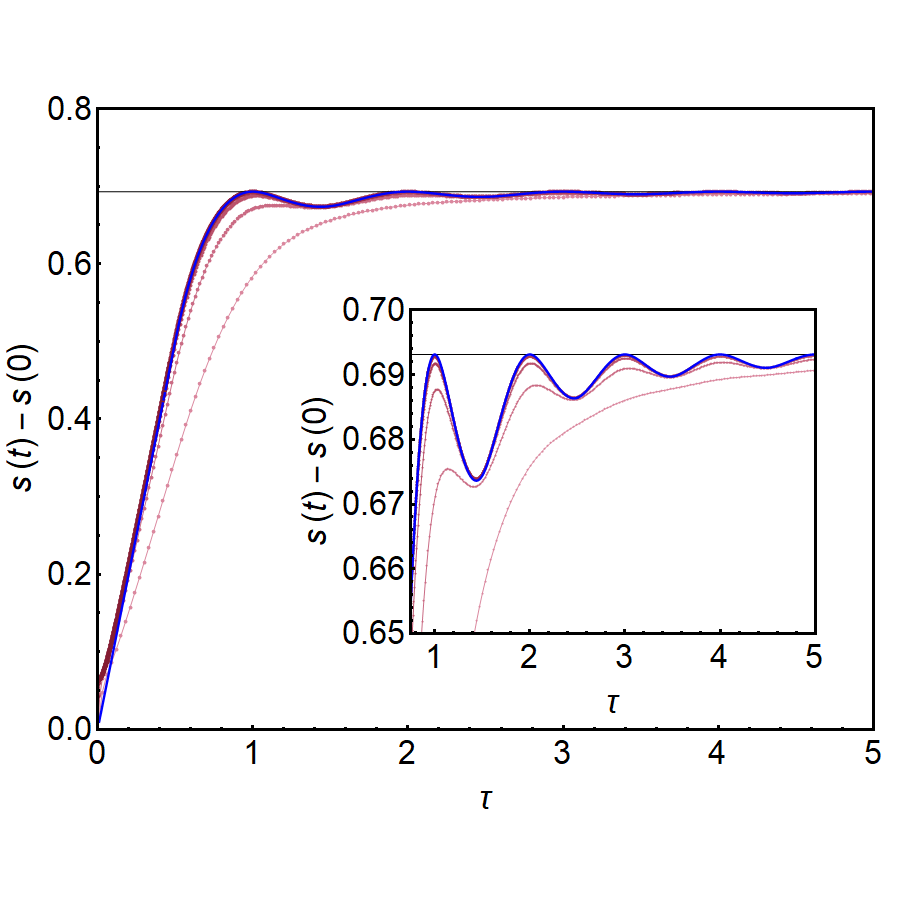}  
\captionof{figure}{Left: Boltzmann entropy for the ideal gas with $N=10^8$ particles, with $n_x=16$  
and $n_v=(4, 8, 16, 32, 64, 128, 256)$.  
The colors intensify as $n_v$ increases. Right: The same data with time rescaled: $\tau = t\Delta v_x/L$;  
$L=1$. The blue solid curve represents the theoretical result for the ideal gas (see Ref. \cite{dhar}) in the limit $\Delta x$, $\Delta v_x \rightarrow 0$. }. 
\label{figi1}  
\end{center}

\subsection{Finite $\lambda$}
In Fig. \ref{figd1} we present the results for $S_B(M_g(t))/N$ for $N=10^5$ hard discs with an areal density of $\eta=10^{-6}$ (mean free path $\lambda\simeq 0.7$) and $n_x=16$ cells along the $x$ direction, with different values of $\Delta v_x$.  For the largest value of $\Delta v_x$ shown, the entropy production is mostly due to the dispersion from the free particle motion, so the behavior is close to that of the ideal gas.  For smaller $\Delta v_x$, however, the entropy production due to the interparticle scattering dominates, and the behavior becomes very different from the ideal gas.  In the left panel of Fig. \ref{figd1}, the behavior becomes almost independent of $\Delta v_x$ for small $\Delta v_x$, since the free particle dispersion then stops giving an important contribution to the entropy production.


\begin{center}
\includegraphics[height=7cm]{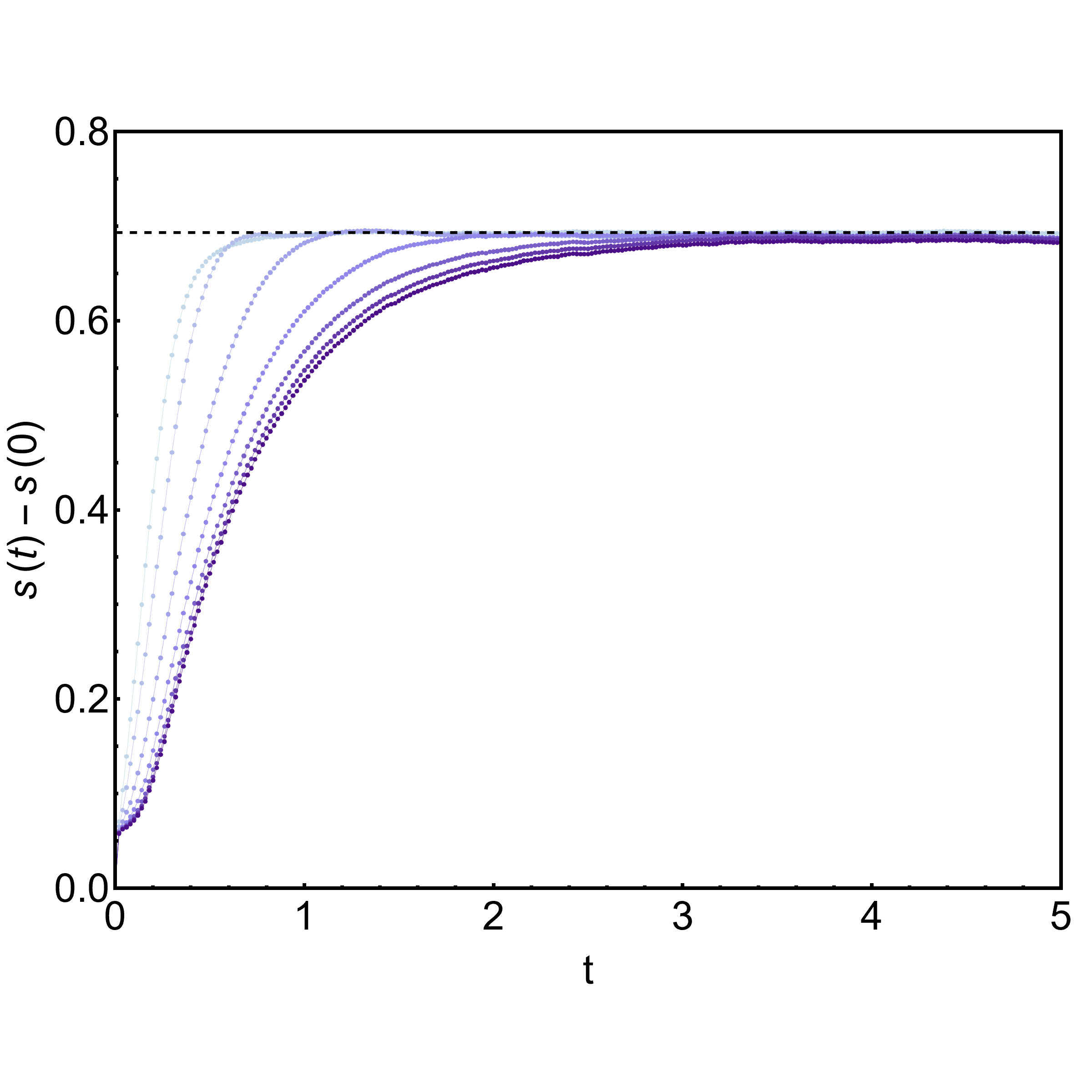}  
\includegraphics[height=7cm]{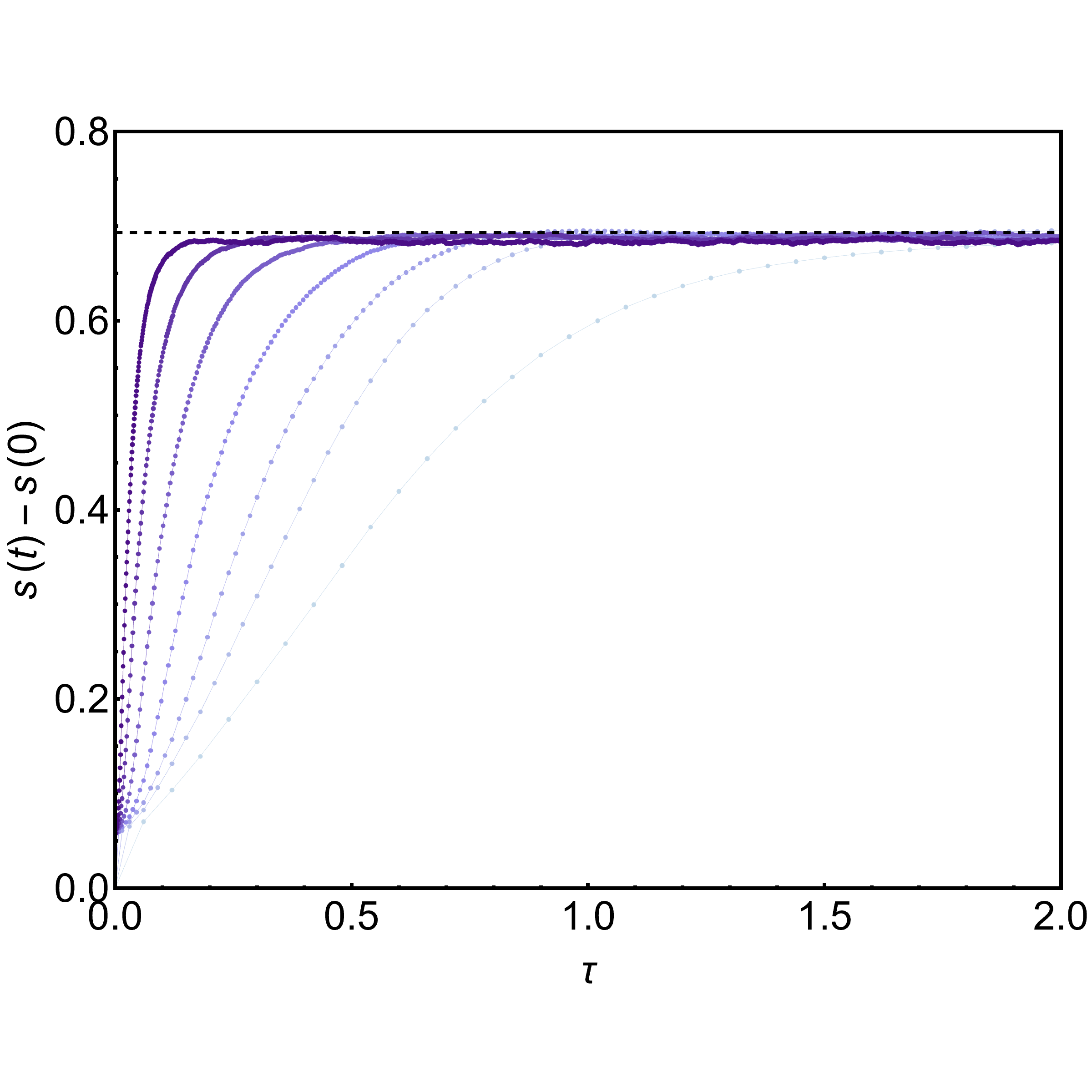}  
\captionof{figure}{Left: Boltzmann entropy for $N=10^5$ particles and $\eta=10^{-6}$ (mean free path $\lambda\simeq 0.7$), with $n_x=16$  and $n_v=(4, 8, 16, 32, 64, 128, 256)$.  
The colors intensify as $n_v$ increases. Right: The same data with time rescaled as was appropriate for the ideal gas: $\tau = t\Delta v_x/L$: 
note that this scaling does not collapse these data.} 
\label{figd1}  
\end{center}

In Fig. \ref{fig00} we show the time evolution of the single-particle phase space distributions in $(x,v_x)$ for three different values of the areal density corresponding to $\lambda\simeq 0.07$ ($\eta=10^{-4}$), $\lambda\simeq 0.7$ ($\eta=10^{-6}$), and the ideal gas ($\eta=0$).  The initial state has the gas confined to occupy only one half of this phase space, namely $0.0<x<0.5$.  For the ideal gas (right panels) the free dynamics of this single-particle marginal distribution is simply $f(x,v_x,t)=f(x-v_x t,v_x,0)$, so it always occupies only one half of the single-particle phase space; this is due to the conservation of the one-particle phase space volume in the absence of interactions.
The initial region 
develops into stripes that become finely spaced along the $v_x$ direction, getting finer as time increases.  {\bf Thus the entropy of the ideal gas does not increase due to occupying a larger fraction of the single-particle phase space, instead it increases due to the stripes that it occupies becoming more finely spaced.}  
For the dilute gas (center), scattering events will  
move some of the particles, eventually one half of them, into the region that the ideal gas does not visit.  As time progresses, this scattering produces entropy and reduces the contrast between the dark and light stripes in the center and left panels of Fig. \ref{fig00}, with the rate of this scattering larger for the larger discs (left).


\begin{center}
\includegraphics[width=5cm]{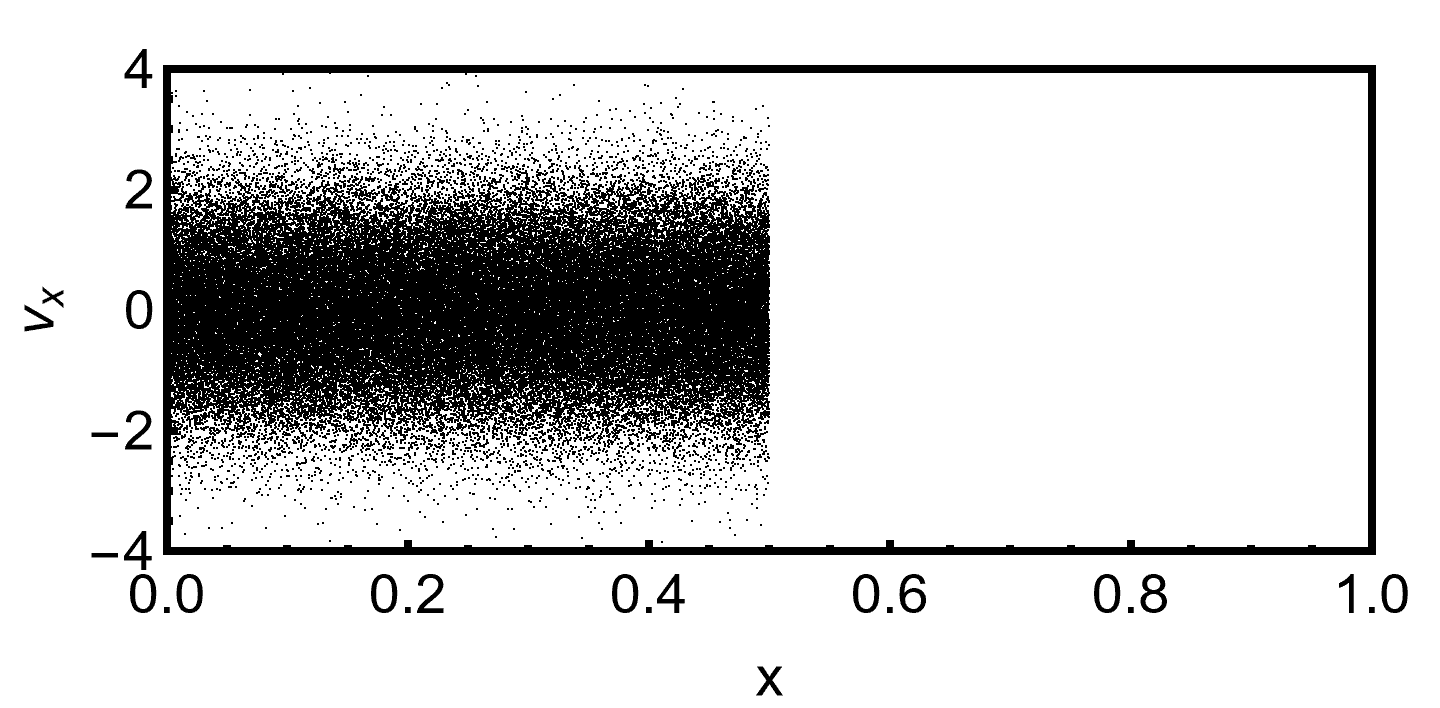} 
\includegraphics[width=5cm]{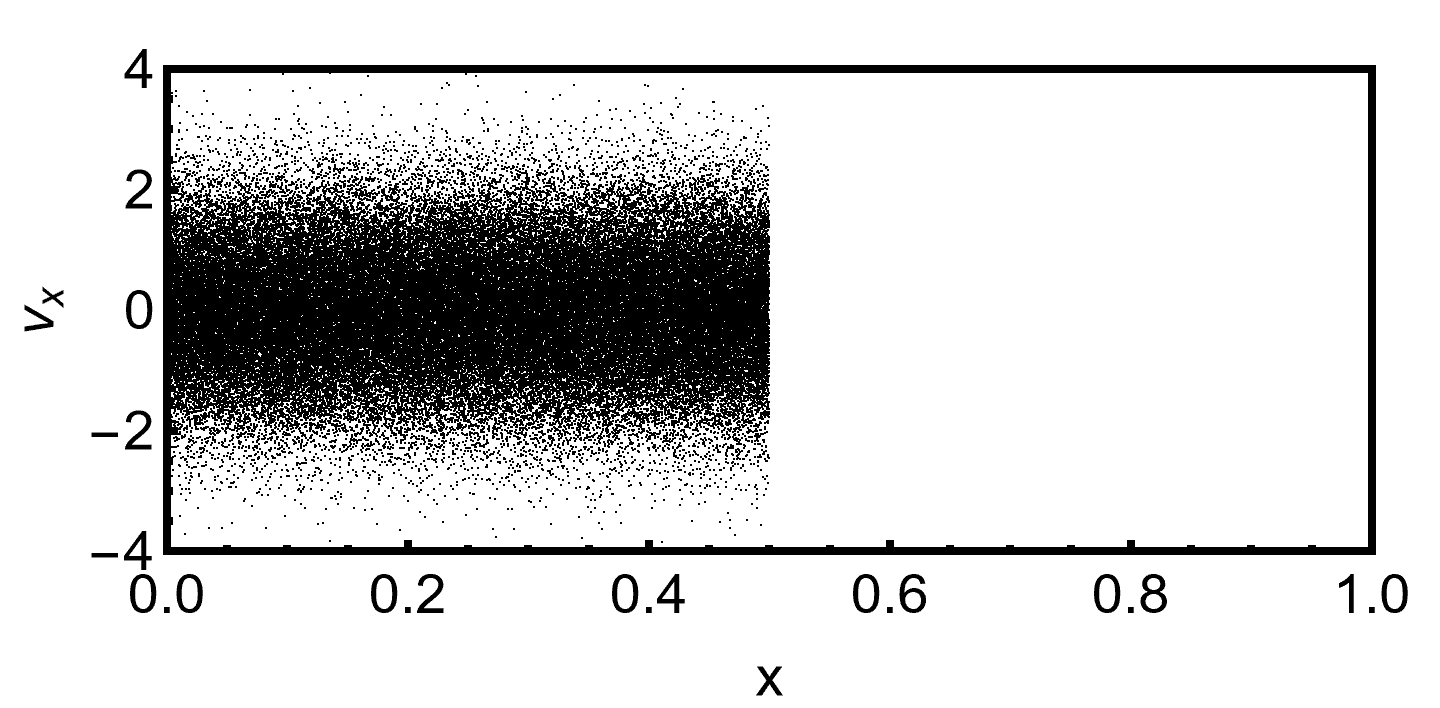}
\includegraphics[width=5cm]{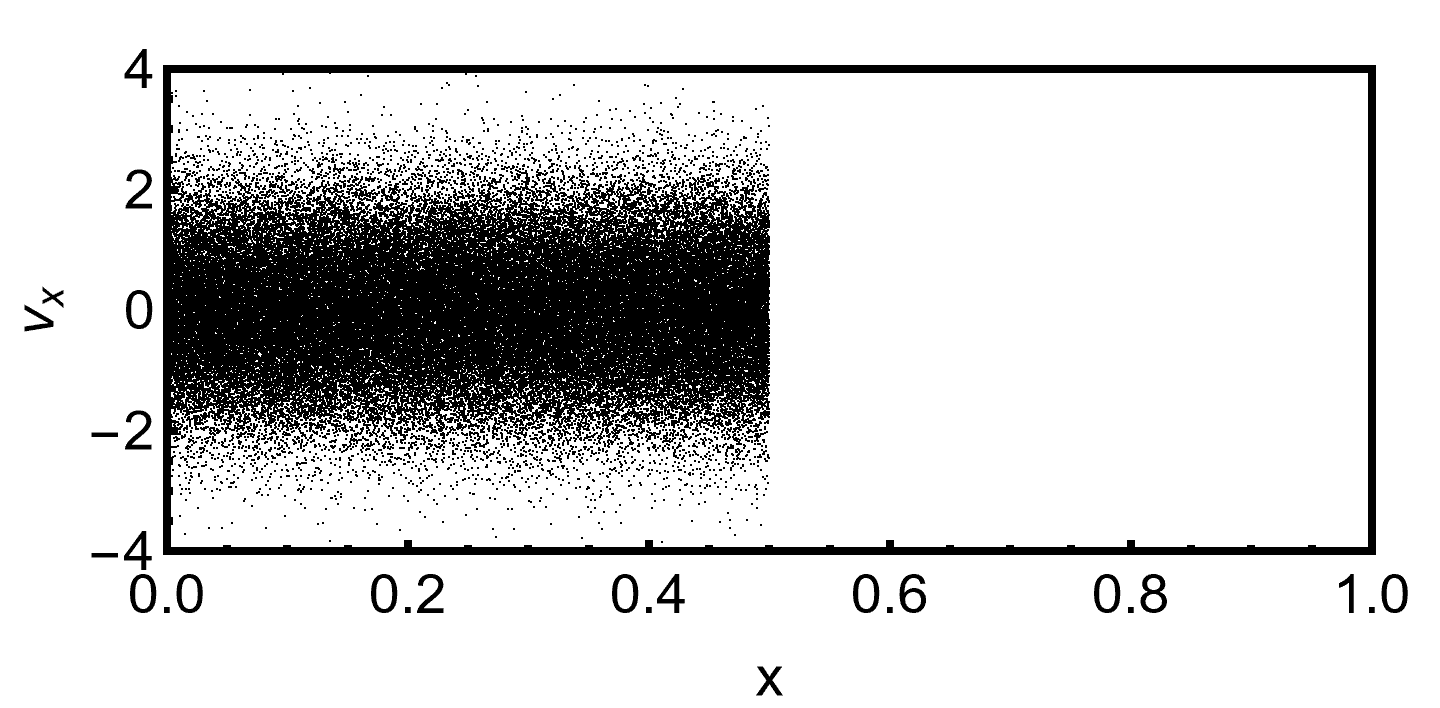}
\vskip -0.1truecm
\includegraphics[width=5cm]{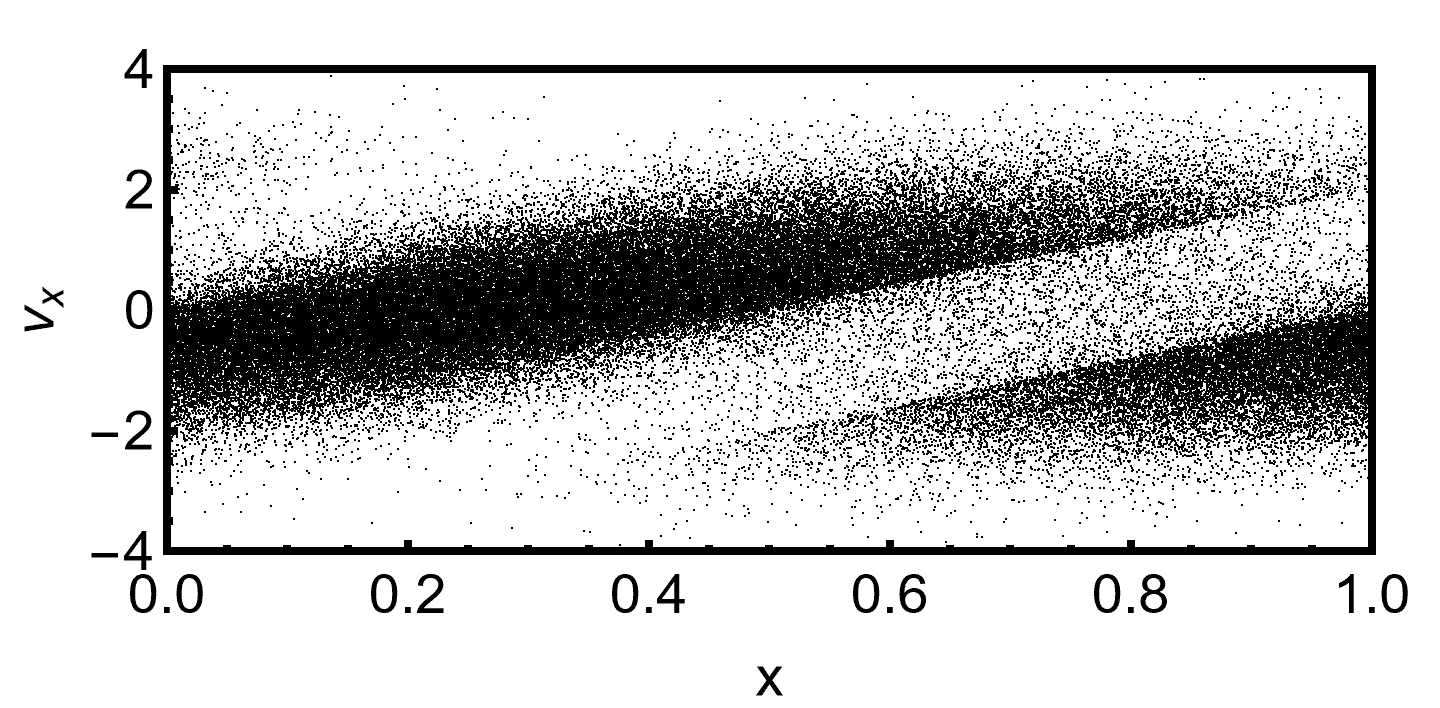} 
\includegraphics[width=5cm]{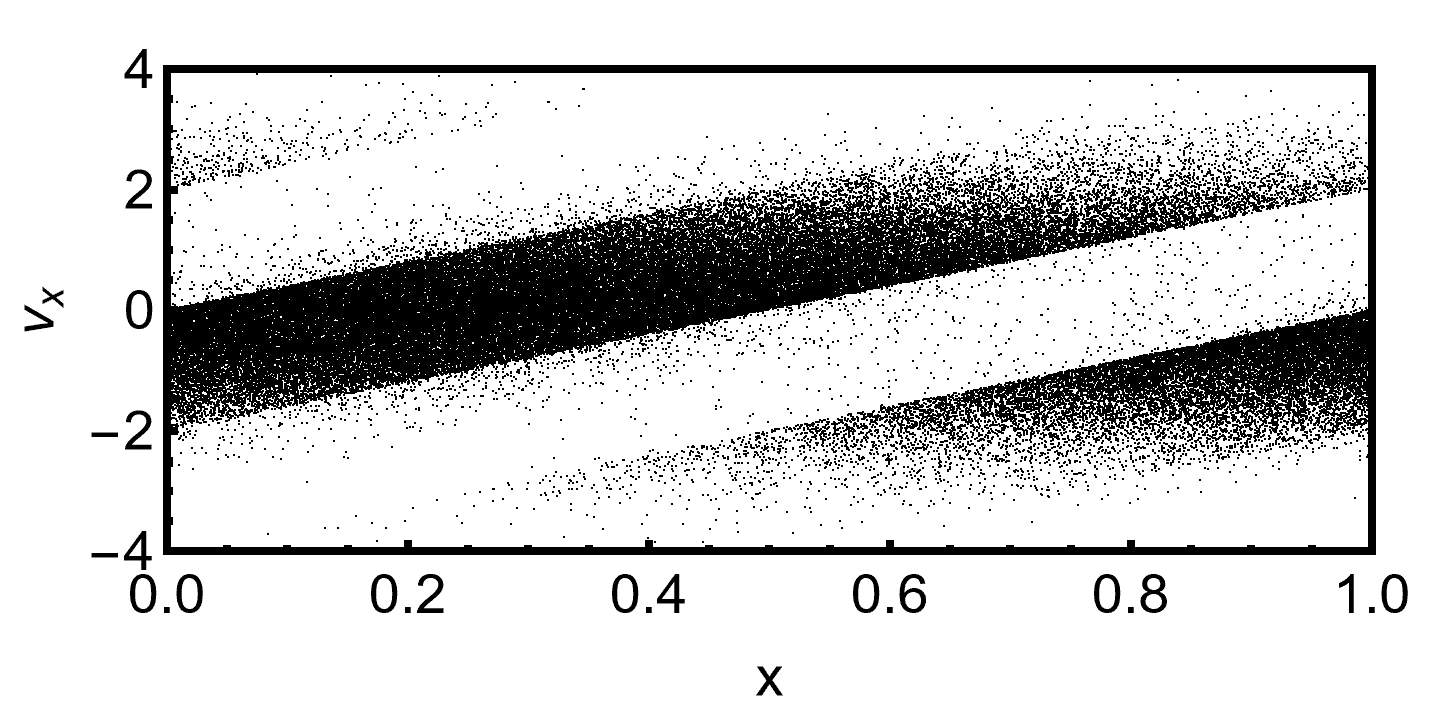}
\includegraphics[width=5cm]{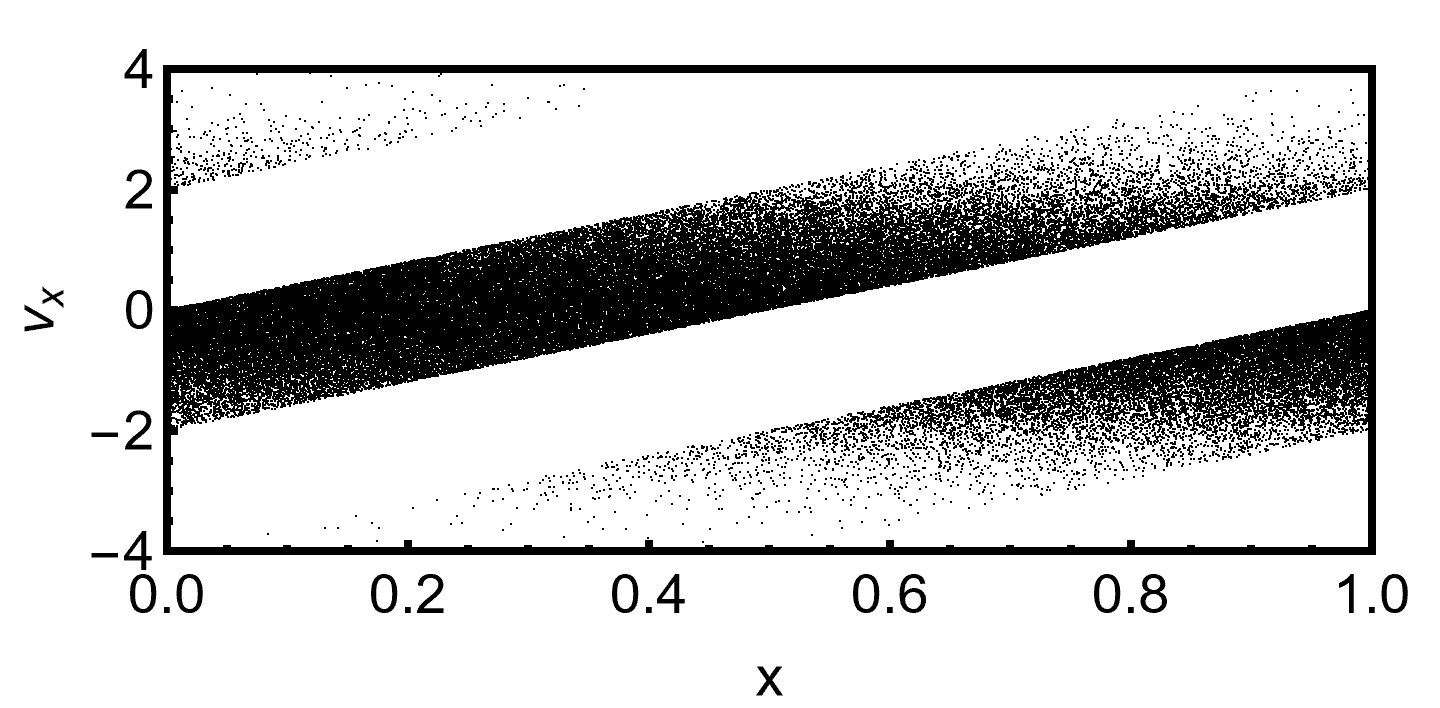}
\vskip -0.1truecm
\includegraphics[width=5cm]{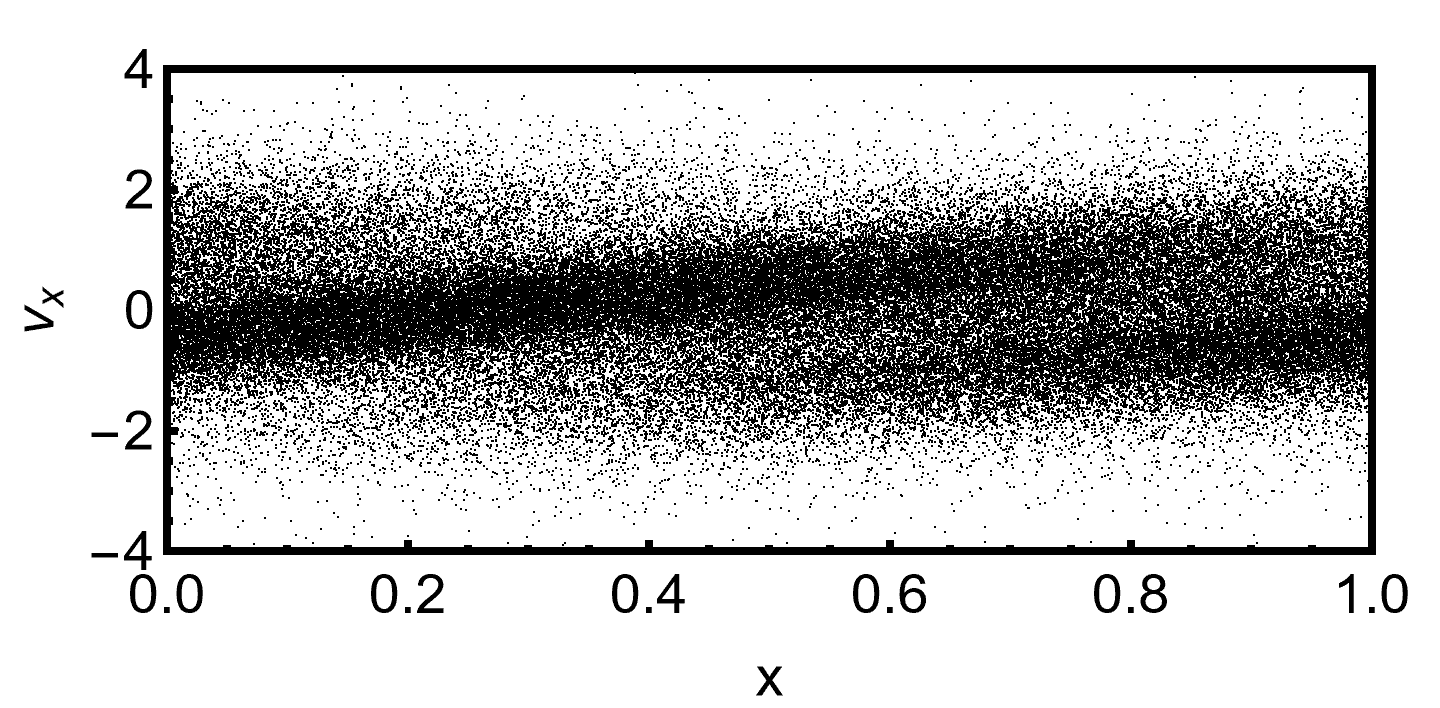} 
\includegraphics[width=5cm]{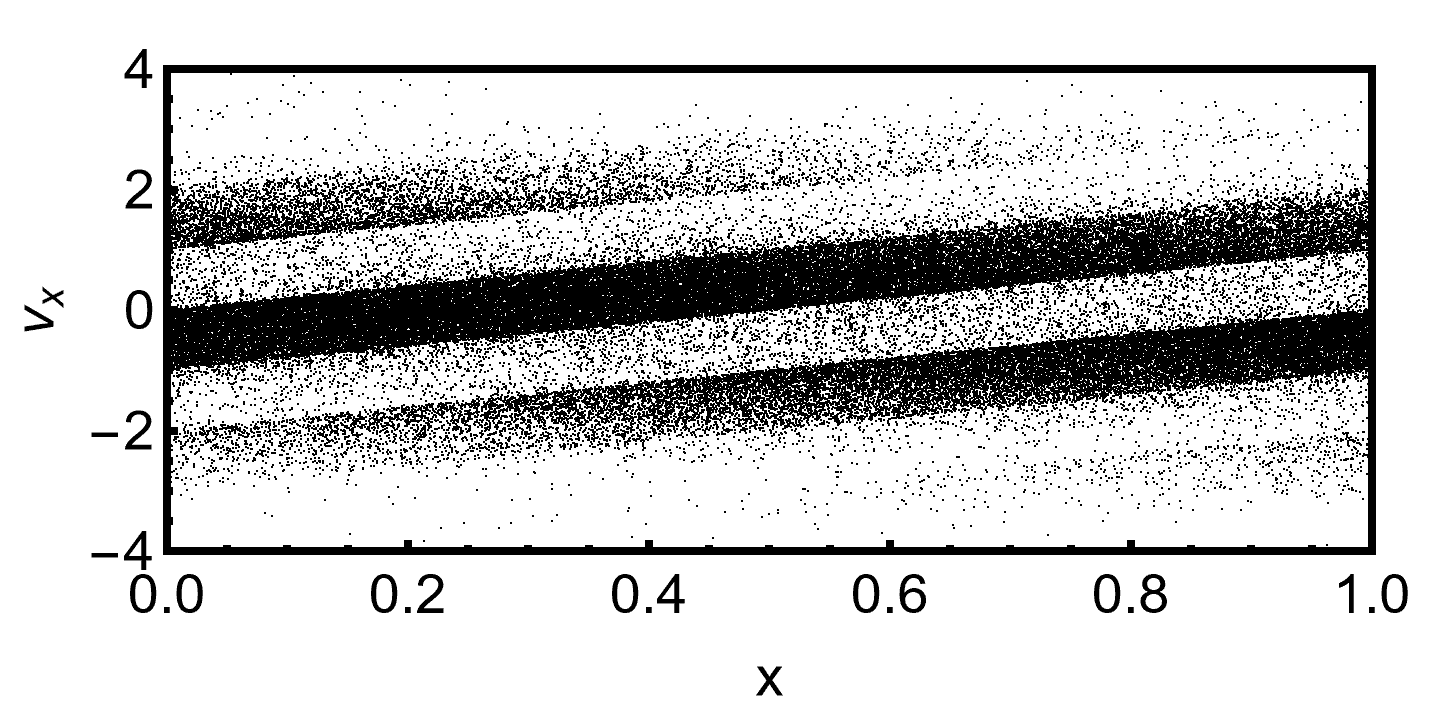}
\includegraphics[width=5cm]{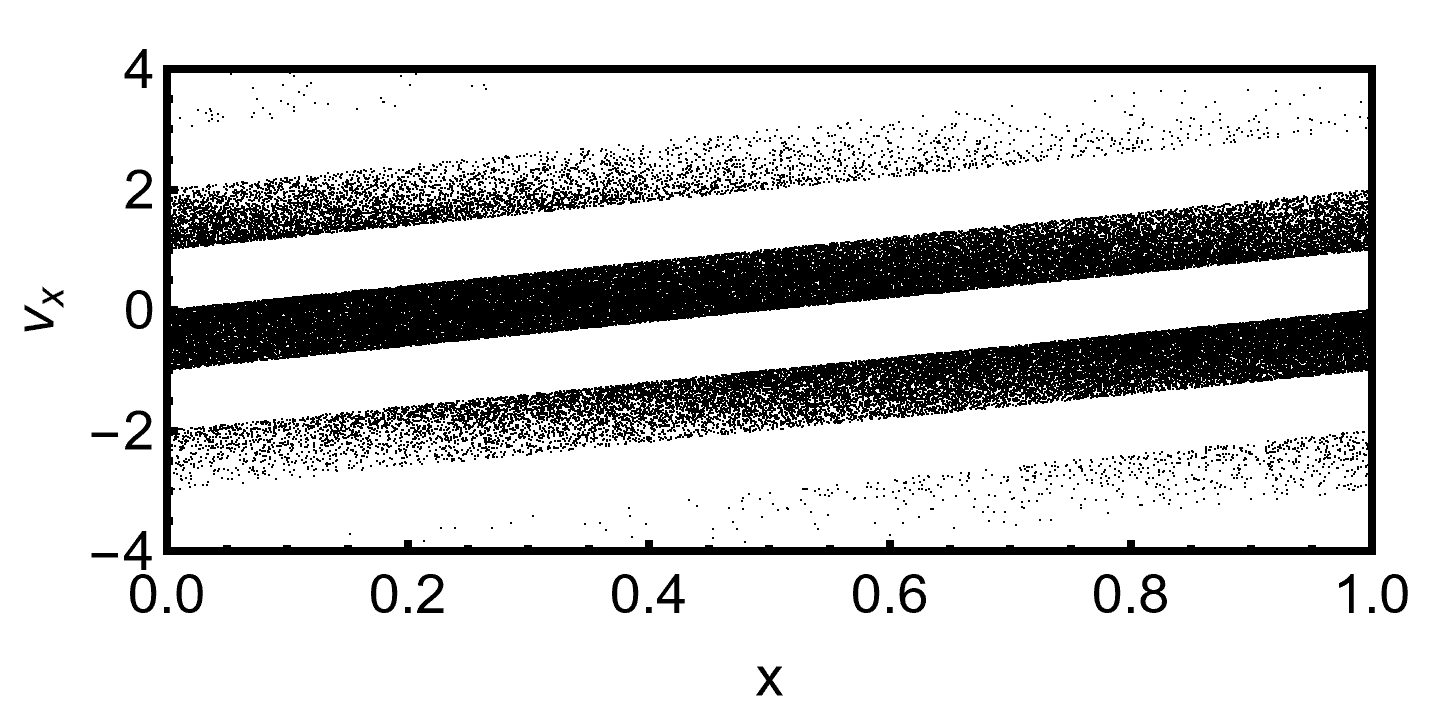}
\vskip -0.1truecm
\includegraphics[width=5cm]{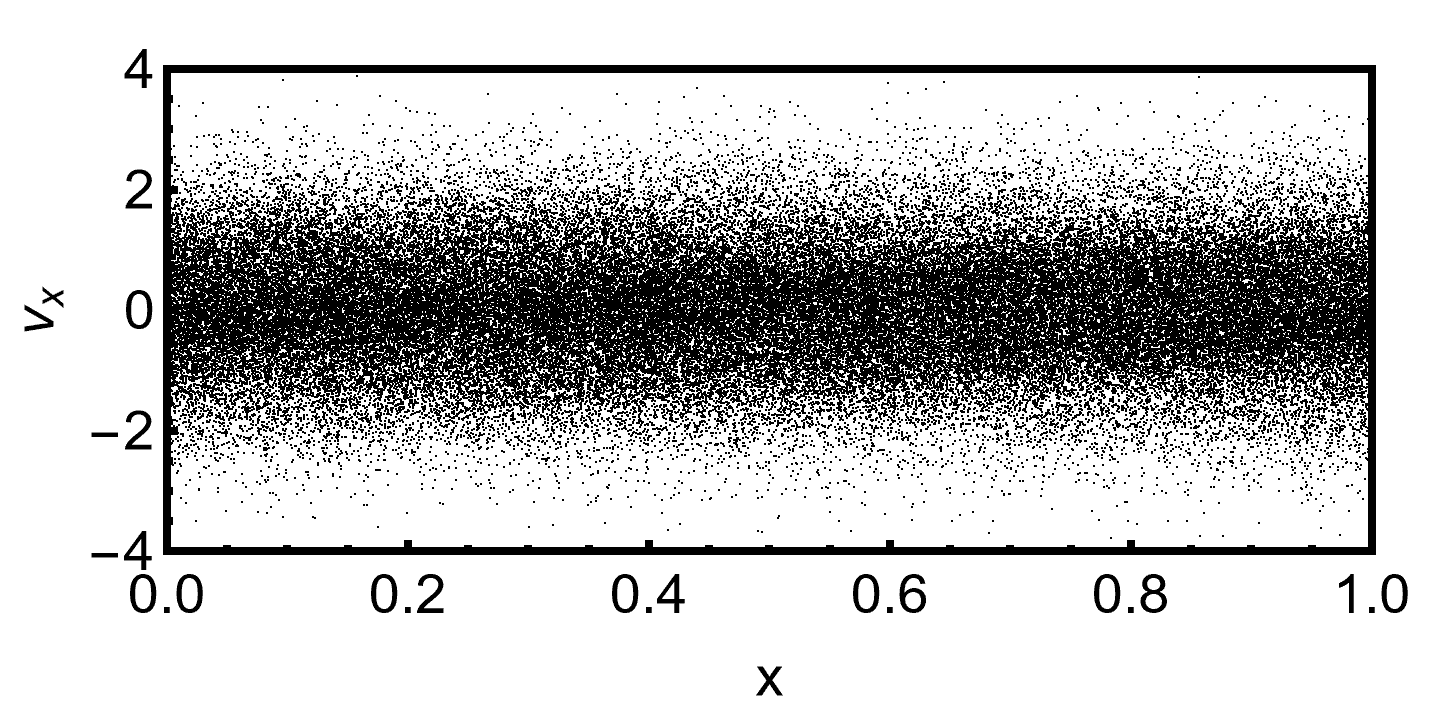} 
\includegraphics[width=5cm]{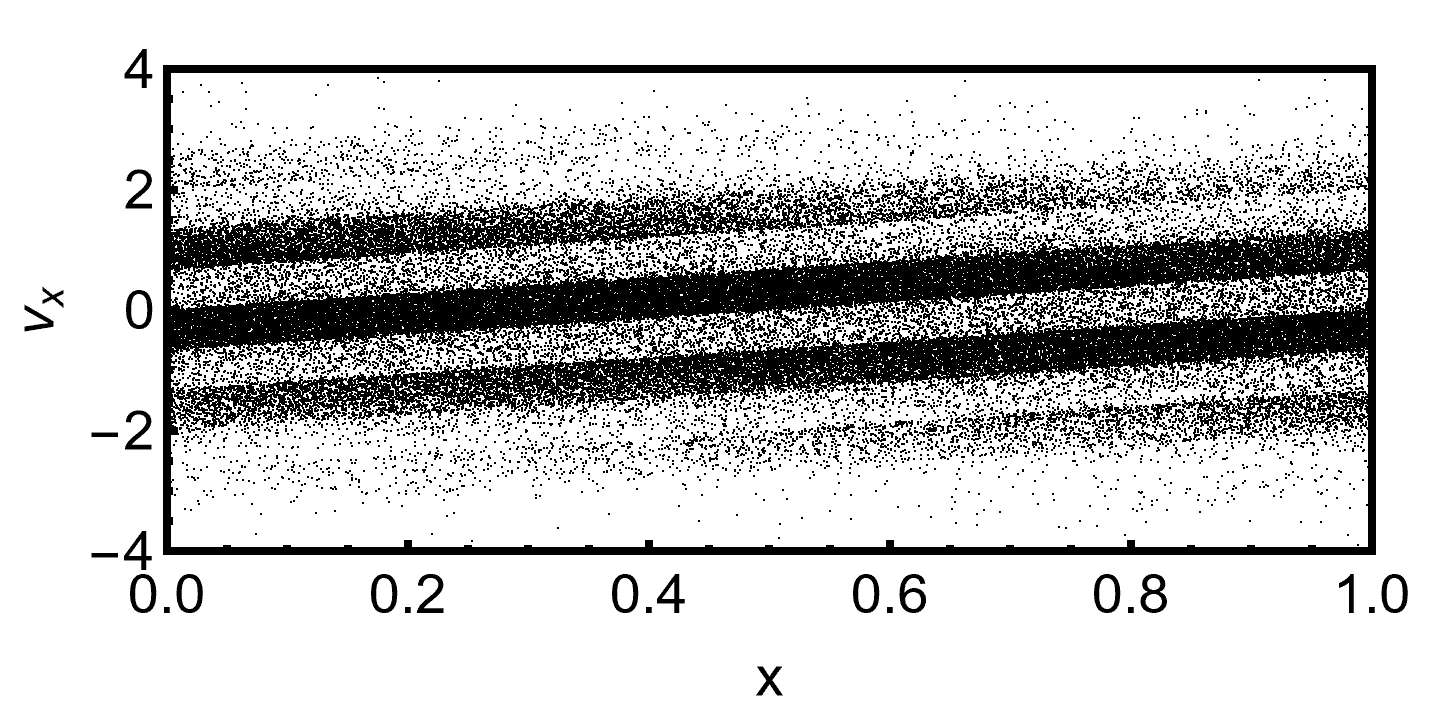}
\includegraphics[width=5cm]{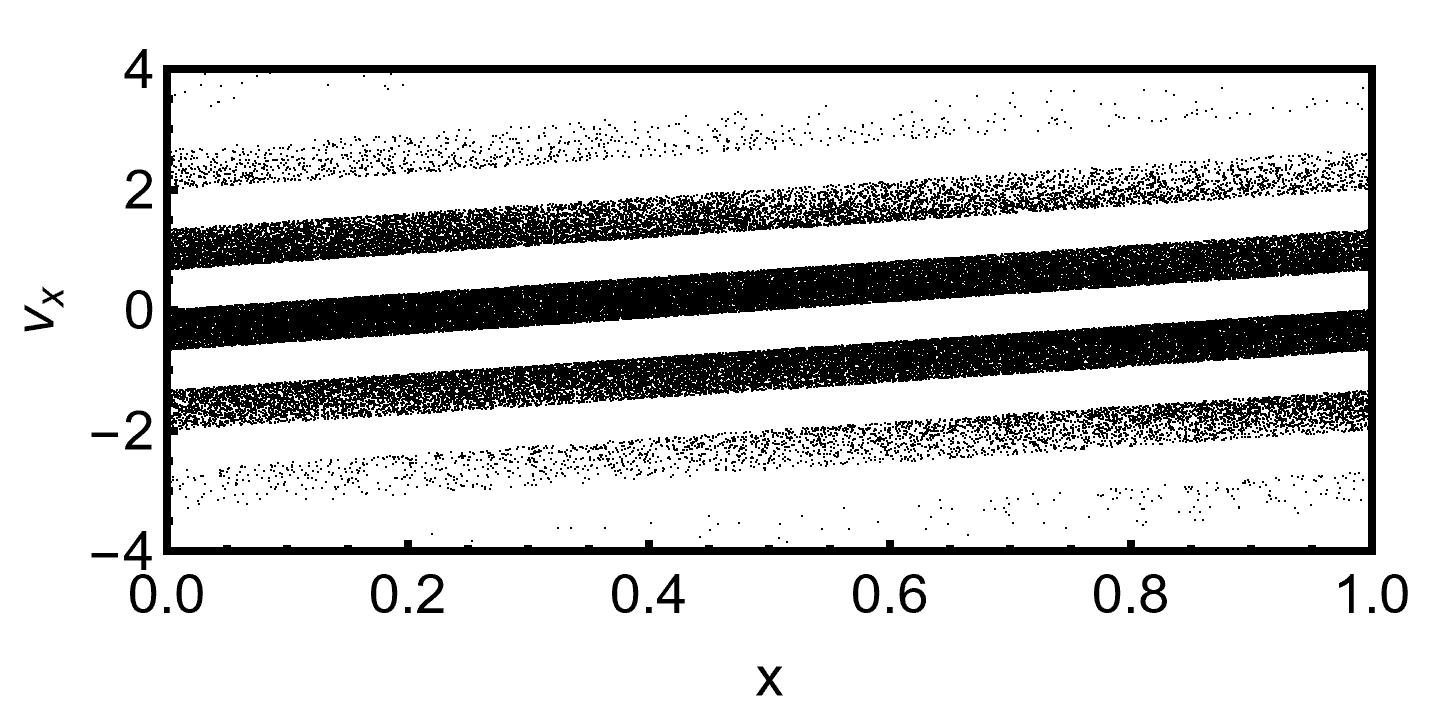}
\vskip -0.1truecm
\includegraphics[width=5cm]{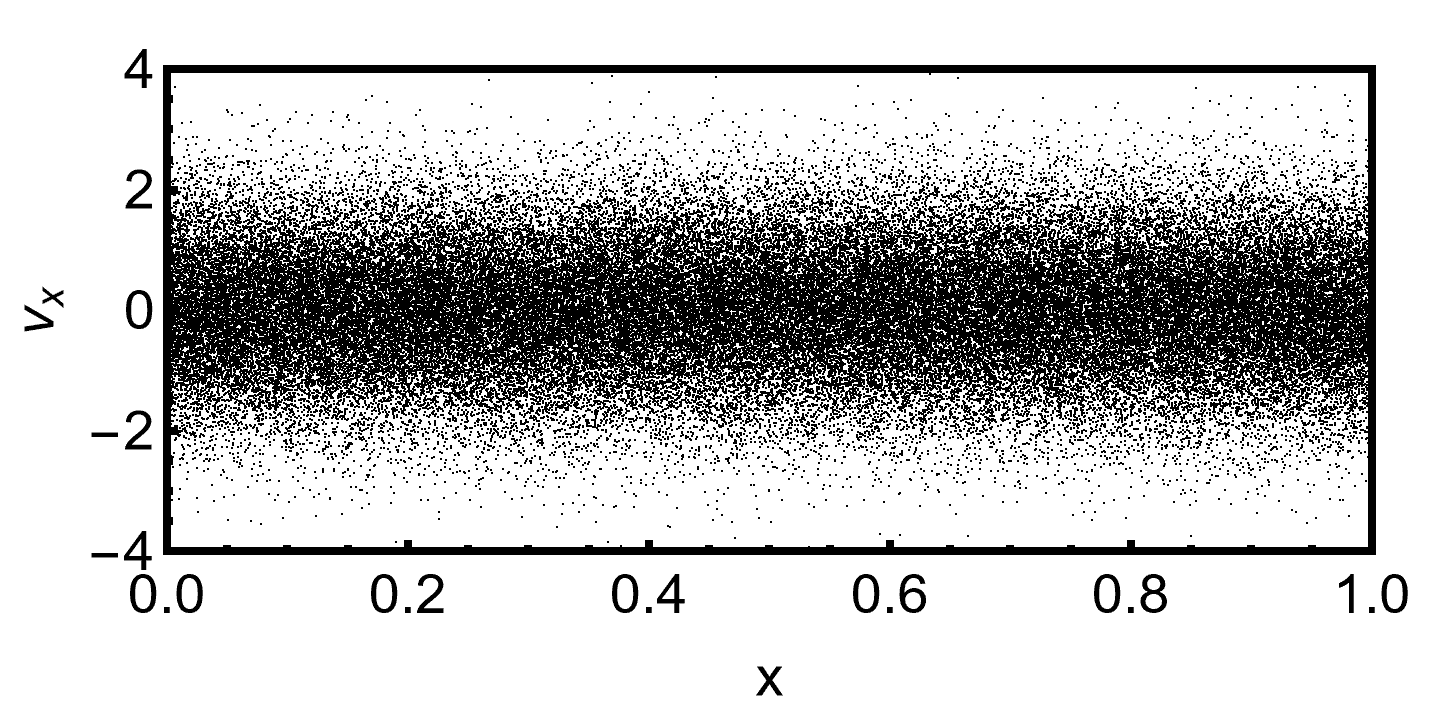} 
\includegraphics[width=5cm]{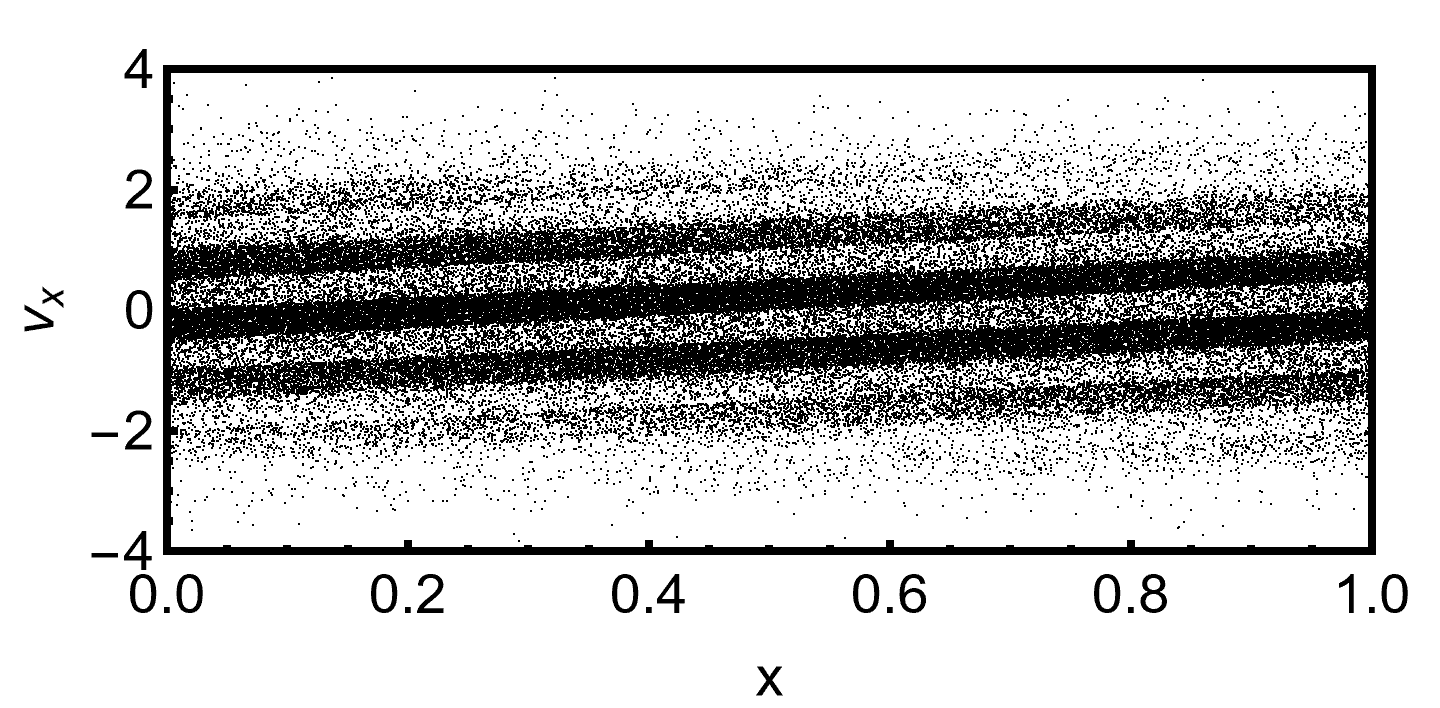}
\includegraphics[width=5cm]{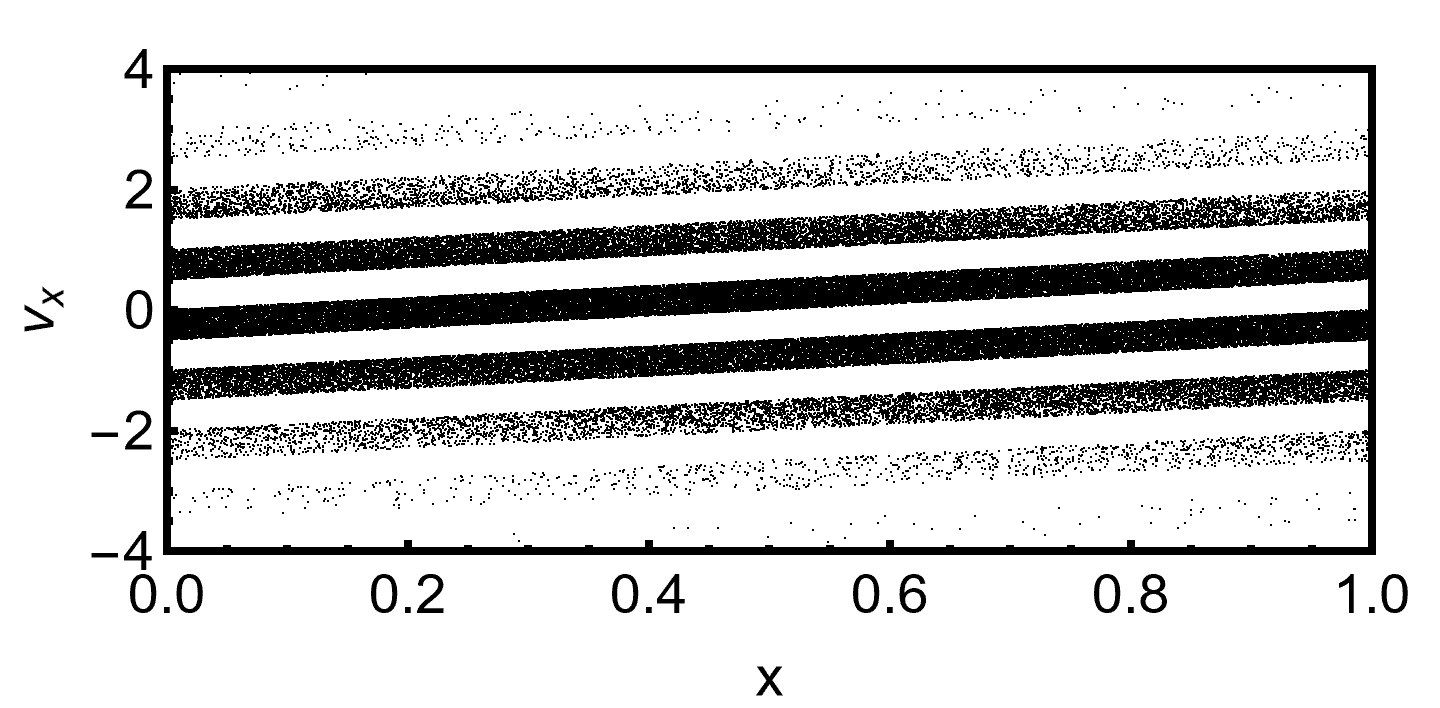}
\captionof{figure}{$(x,v_x)$-phase space evolution of a system with $N=10^5$ discs with areal densities $\eta=10^{-4}$, mean free path $\lambda\simeq 0.07$, (left column); $\eta=10^{-6}$, mean free path $\lambda\simeq 0.7$, (center column); and the ideal gas, $\lambda^{-1}=0$ (right column) for times (from top to bottom): $t=0.0$,  $0.25$,  $0.50$,  $0.75$ and $1.0$.} 
\label{fig00}  
\end{center}

\section{Analysis of entropy production for dilute hard discs}

To obtain a theoretical description of the entropy evolution shown in Fig. \ref{figd1} we follow the analysis in Ref \cite{dhar}.  
Instead of looking at a single microstate we consider $F(x,v_x,t)$, the marginal one-particle distribution of the time-evolved $N$-particle distribution following the lifting of the constraint at time $t=0$~:
\begin{equation}
F(x,v_x,t)=\sum_{i=1}^N \langle\delta(x_i(t)-x)\delta(v_{x,i}(t)-v_x)\rangle ~.
\end{equation}
Note that $F$ is normalized so $\int dx\, dv_x F = N$.  We take the initial distribution, before the lifting of the constraint, to be the $N$-particle 
canonical Gibbs distribution in the smaller box, $L_x=1/2$, $L_y=1$. We then coarse grain the time-evolved $F$ over each cell $\Delta_\alpha$:
\begin{equation}
F_\alpha(t)=\frac{1}{\Delta}\int_{\Delta_\alpha}\, dx\, dv_x\, F(x,v_x,t),\label{en7}
\end{equation}
where $\Delta=|\Delta_\alpha|$.  

Using the law of large numbers we expect that the average number of particles in $\Delta_\alpha$ is close to that of single (typical) realizations.  Hence the coarse grained entropy per particle of the $x$ and $v_x$ degrees of freedom
\begin{equation}
s^F_\Delta=-\frac{\Delta}{N} \sum_\alpha F_\alpha \log F_\alpha 
\label{en8}
\end{equation}
will be close to the Boltzmann entropy of a typical microstate as given in Eq. (\ref{eq:entz}).  This was verified in Ref. \cite{dhar} for the ideal gas where $F(x,v_x,t)$ is simply the solution of Eq. (\ref{eq:free}): $F_{ideal}(x,v_x,t)=F(x-v_x t,v,0)$.

We will now argue that $F(x,v_x,t)$  
has a rather simple approximate form when the initial velocity distribution is Maxwellian and we are in the dilute regime $\lambda\gtrsim L$.  We show that this form gives a good approximation to the simulation results for the entropy.   

In the dilute limit there is an early time regime $L/v_{th}\ll t \ll \lambda/v_{th}$ where the gas has spatially expanded to uniformly fill the larger volume.  Very little scattering happens during (or before) this early time regime, so this early time dynamics is well approximated by the ideal gas behavior $F(x,v_x,t)=F(x-v_xt,v_x,0)$.  Since $tv_{th}\gg L$, typical particles have travelled many times the distance $L$, and thus the single-particle distribution has become very finely ``striped'', with the width in $v_x$ of the stripes being $\sim L/t$, which is very narrow compared to the thermal speed $v_{th}$.  The velocity distribution when averaged over position remains Maxwellian, and the position distribution when averaged over velocity is uniform.  Thus the way in which the single-particle distribution is out of thermal equilibrium in the larger box is that it has detailed correlations between $x$ and $v_x$ in the form of these fine stripes along $v_x$.  If the distribution $F(x,v_x,t)$ is coarse-grained in $v_x$ on a scale much larger than these stripes but small compared to $v_{th}$ then it becomes the equilibrium Maxwell distribution.  

Thus, in the dilute limit and in this time regime $L/v_{th}\ll t\ll \lambda/v_{th}$, the single-particle distribution can be written in the form:
\begin{equation}
F(x,v_x,t)=\frac{N}{L}g_T(v_x)+\delta F(x,v_x,t)\quad, 
\end{equation}
where $g_{T}(v_x)$ is the normalized equilibrium Maxwell velocity distribution, and $\delta F(x,v_x,t)$ gives the finely striped correlations between $x$ and $v_x$ that one can see developing with time in Fig. \ref{fig00} and that will vanish ($\delta F\rightarrow 0$) at longer times due to the scattering.  
In the ideal gas, and thus also in this time regime before significant scattering has happened, we have
\begin{equation}
\delta F(x,v_x,t)=F(x-v_x t,v_x,0)-\frac{N}{L}g_T(v_x)~. 
\label{eq:ideal}
\end{equation}
 
For times $t\gg L/v_{th}$, the particles will scatter at a rate, $\Gamma(|\bf{v}|)$, which, since the density is then spatially uniform and the velocity distributions only differ from Maxwellians by the very fine ``striping'', does not depend significantly on $x$ or $t$.  The impact parameter at the collision is approximately uniformly distributed over the width of the particles, which results in the post-scattering velocity distribution being a smooth function of $v_x$, with none of the ``striping'' present in $\delta F$, and the contributions from initial velocities that are very near to each other but have different signs of $\delta F$ will cancel.  Thus the scattering only produces losses from the $\delta F$ part of the distribution, while for the equilibrium (Maxwellian) part of the distribution, the losses and gains are equal and opposite, as they must be at equilibrium.  Since the scattering rate also depends on $v_y$, this will introduce a dependence of $\delta F$ on $v_y$.  But, over the bulk of the Maxwell distribution the dependence of the scattering rate on $|\bf{v}|$ is small (well under a factor of two), so for simplicity we will make the approximation that it is a constant: $\Gamma(|{\bf v}|)=\Gamma$.

In this approximation, $\delta F(x,v_x,t)$ is reduced from the ideal gas form (\ref{eq:ideal}) by a factor of $\exp(-\Gamma t)$, so the behavior for $t\gg L/v_{th}$ in the dilute limit is well approximated by  
\begin{equation}
F(x,v_x,t)=e^{-\Gamma t} F(x-v_xt,v_x,0)+\left(1-e^{-\Gamma t}\right)\frac{N}{L}g_T(v_x)~.
\label{eq:solrta}
\end{equation}
We note that (\ref{eq:solrta}) is the solution to the linearized relaxation time approximation to the BE: 
\begin{equation}
\frac{\partial F(x,v_x,t)}{\partial t} + v_x \frac{\partial F(x,v_x,t)}{\partial x} = \Gamma\left[\frac{N}{L}g_T(v_x)-F(x,v_x,t)\right]~.
\label{eq:rta}
\end{equation}
In fact one can think of (\ref{eq:solrta}) as a simple, perhaps the simplest, interpolation between the short and long time behavior of $F$. We shall use it now to compute $s^F_\Delta$ in (\ref{en8}).

The form of (\ref{eq:rta})  
clearly delineates the two distinct processes contributing to entropy production in this dilute gas. Collisions occur at a rate of $\Gamma$ per particle, leading to entropy production per particle and an approach to equilibrium at a rate $\sim \Gamma$, which is insensitive to the choice of cell size $|\Delta_\alpha|$. On the other hand, the free motion for a time $t$ of particles initially in a cell with width $\Delta v_x$ spreads those particles over an $x$ range $\sim t\Delta v_x$, resulting in entropy production when this length scale becomes larger than the initial length scale $L$ of the spatial inhomogeneity of the density. Thus, the free motion contributes to entropy production per particle at a rate $\sim \Delta v_x/L$, linearly dependent on our choice of $\Delta v_x$. The entropy production due to free motion becomes significant when the scattering is weak, such that $\Gamma \lesssim \Delta v_x/l$, where $l$ is the length scale of any initial density inhomogeneities.

We can estimate the scattering rate $\Gamma$ based on the evolution of a typical single configuration. At each time $t$, we measure the fraction of particles, denoted as $\nu(t)$, whose noninteracting backward evolution over a time $t$ leads them back to the initial half box:
\begin{equation}
\nu(t)=\frac{1}{N}\sum_{i=1}^N\chi\left(0<(x_i(t)-v_{x,i}t)<L/2\right)~,
\end{equation}
where $\chi(\cdot)$ is the indicator function.  Assuming that after any collision the noninteracting backward time evolution is equally likely to put the particle in either of the initial half boxes, $\nu(t)$ should be the sum of the fraction of particles that had not had any collision by time $t$, denoted as $p(t)$, and half of the fraction of particles that collided, $(1-p(t))/2$. Consequently, we have the relationship $p(t) = 2\nu(t) - 1$.  

We display the behavior of this $p(t)$ for various $\eta$ values in the left panel of Fig. \ref{fignu}. From the data, we take $\Gamma$ used in (\ref{eq:solrta}) as the fitted slope of $-\log(2\nu(t)-1)$. The fitted values of $\Gamma$ are plotted in the right panel of  Fig. \ref{fignu} vs. their corresponding $\sqrt{N\eta}$ to demonstrate their linear relationship, as expected, because $\Gamma\propto\lambda^{-1}\propto\sqrt{N\eta}$.
\begin{center}
\includegraphics[height=7cm]{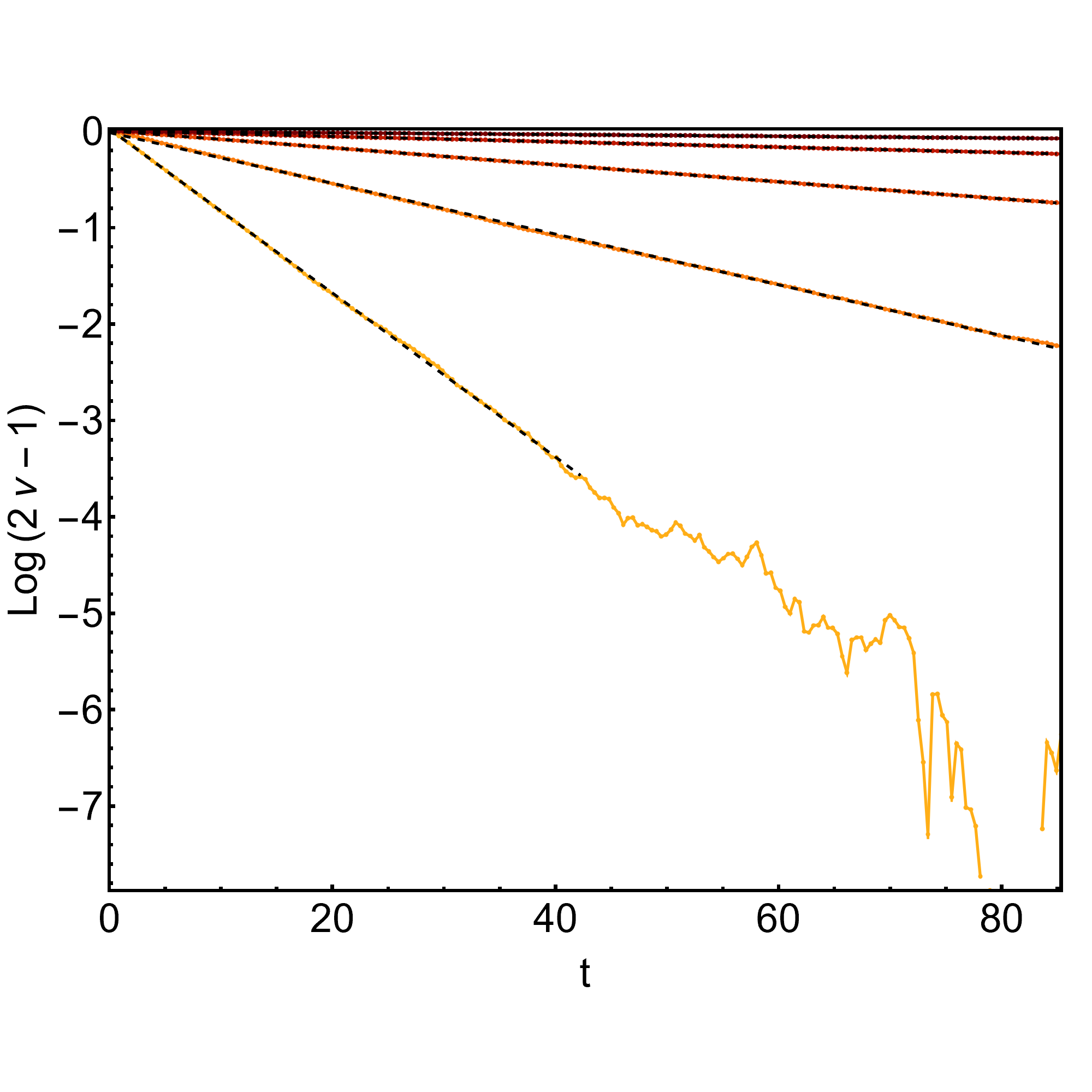}  
\includegraphics[height=7cm]{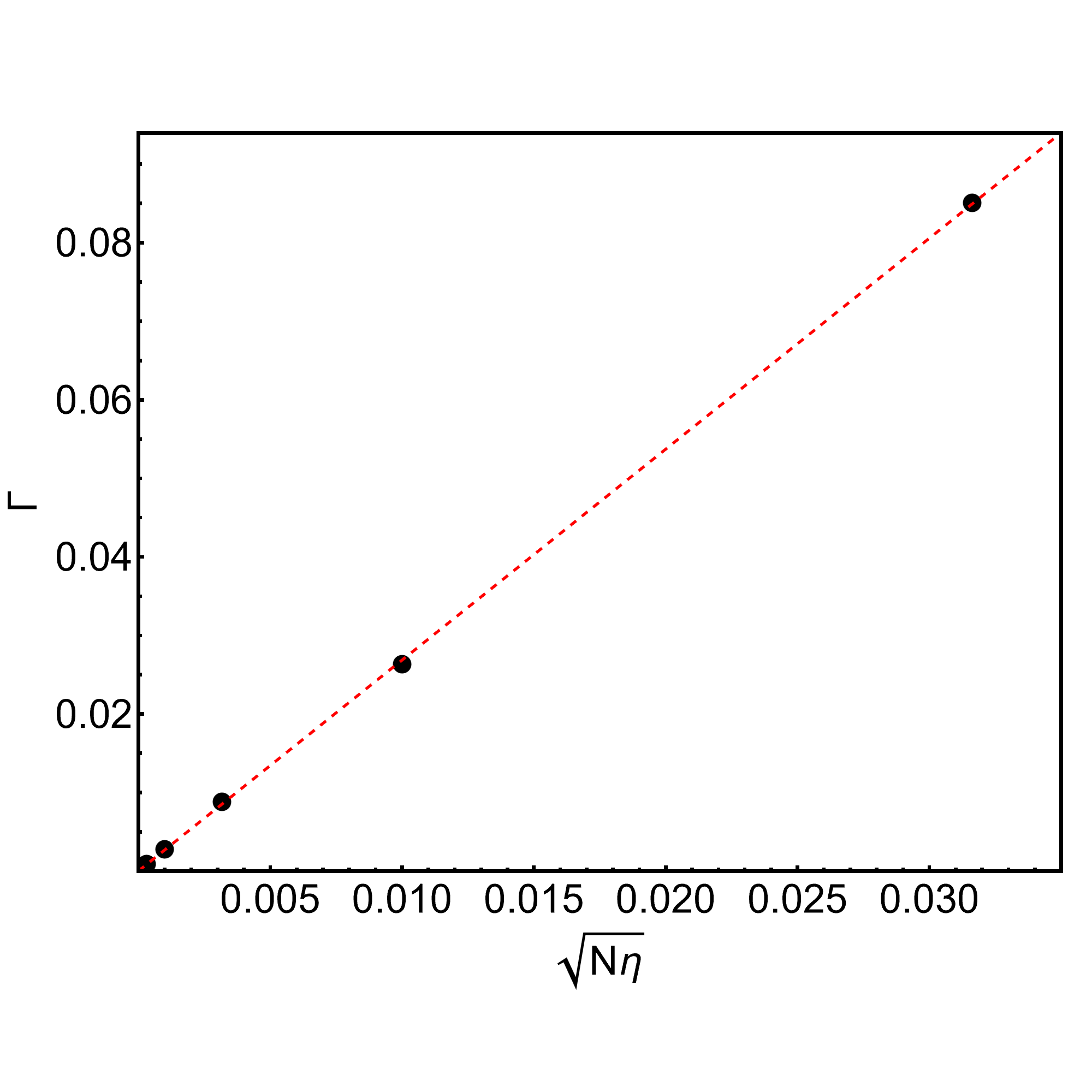}  
\captionof{figure}{Left: $log(2\nu-1)$ vs. $t$ and different $\eta$'s: $10^{-8}$, $10^{-9}$, $\ldots$, $10^{-12}$  that correspond to the mean free paths  $\lambda\simeq 7.01$, $22.2$, $70.1$, $222$ and $701$, respectively.  The colors intensify as $\eta$ decreases. Black dashed lines are linear fits to the data whose slope is $\Gamma(\eta)$. Right: measured slopes $\Gamma(\eta)$ vs $\sqrt{N\eta}$. Red dashed line is the linear fit: $\Gamma(\eta)=2.69(3) \sqrt{N\eta}/L$,~$L=1$.} 
\label{fignu}  
\end{center}

We use equation (\ref{eq:solrta}) with the fitted value of $\Gamma$ to obtain the coarse-grained distribution $F_\alpha$ as given by equation (\ref{en7}). We then calculate the entropy, $s_\Delta^F$, given in equation (\ref{en8}). The details and some properties are presented in the Appendix. The crucial observation is that the functional form of the scaled time-dependence of the entropy in the dilute regime is set by the ratio, $C=\Gamma L/\Delta v_x$, between the rates of entropy production due to the scattering and that due to the dispersive free motion.
Therefore, we expect that any sequence $(\eta,\Delta v_x)$ having a given fixed value of $C\propto \sqrt{\eta}/\Delta v_x$ will tend to the same scaled time-dependence of the entropy as $\Delta v_x$ and $\eta$ go to zero.
We have done a set of simulations with $N=10^5$ particles and $L=1$ to check this limiting behavior.  
We take the values of $C$ corresponding to four different values of $\eta_0$ with $n_v=256$ velocity cells, see Fig. \ref{fit1}. For each $\eta_0$ we simulate the sequence of values:
$(\Delta v_x,\eta)=(12/{n_v},\eta_0 (256/n_v)^2$ with $n_v=16, 32, 64, 128, 256$. In each case we evolve the system for a real time $t$ such that we get to scaled time $\tau=t\Delta v_x/L=4$.

We show in figure \ref{fit1} the entropies numerically obtained and compared with the theoretical solution.  We see that in each case the behavior is consistent with converging to the theoretical curves as $\Delta v_x$ and $\eta$ are decreased towards zero for that fixed $C$.

\begin{center}
\includegraphics[width=7cm]{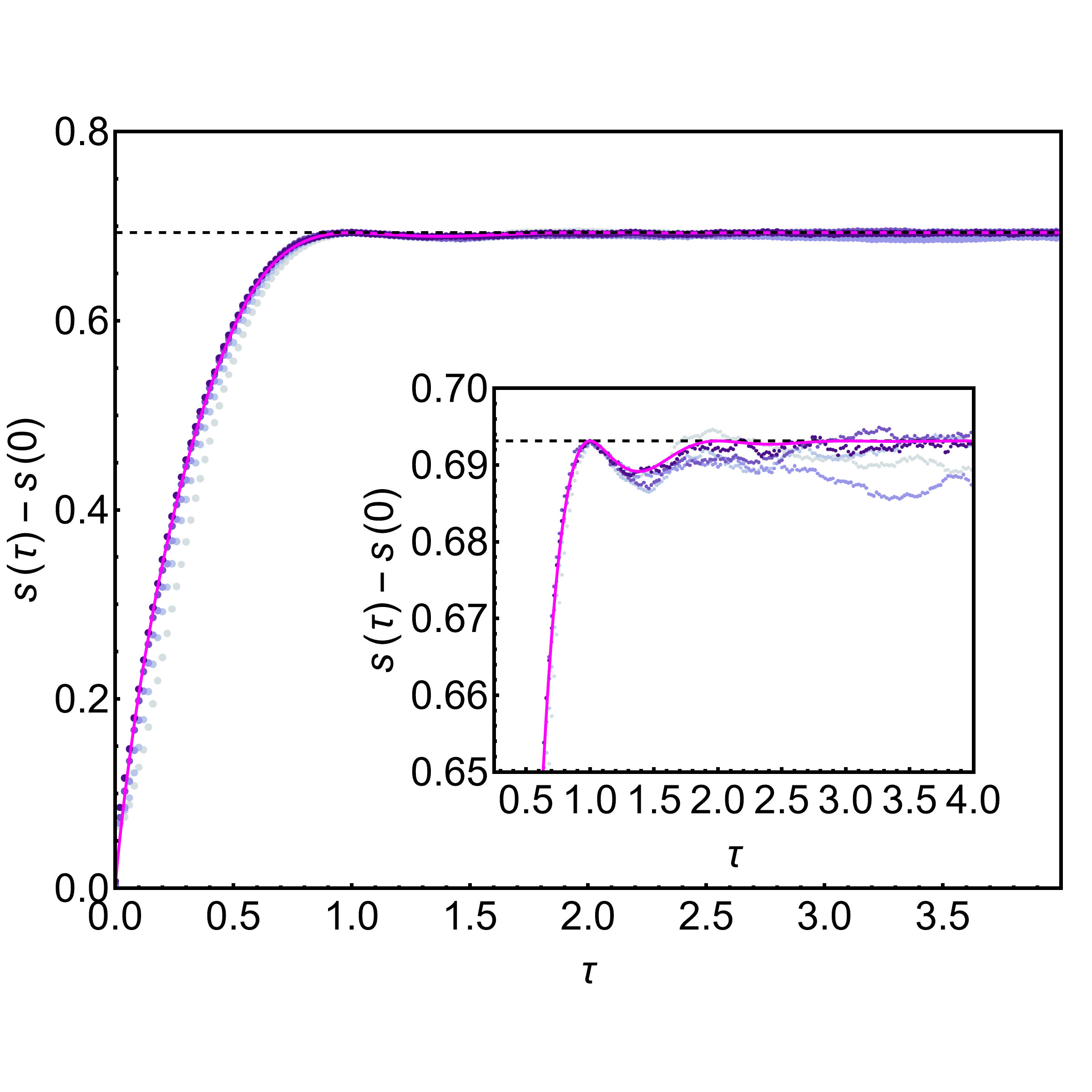}  
\includegraphics[width=7cm]{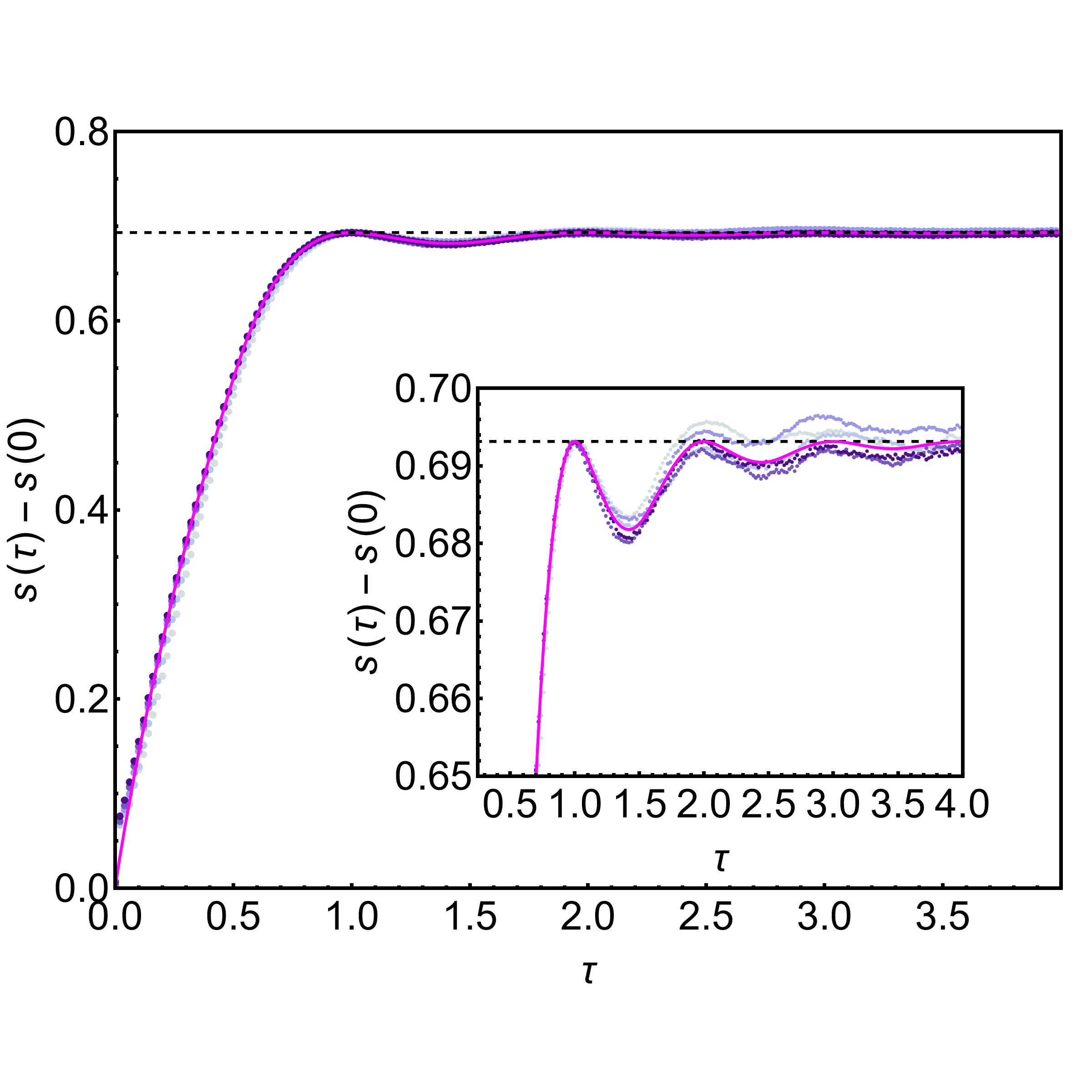}  
\includegraphics[width=7cm]{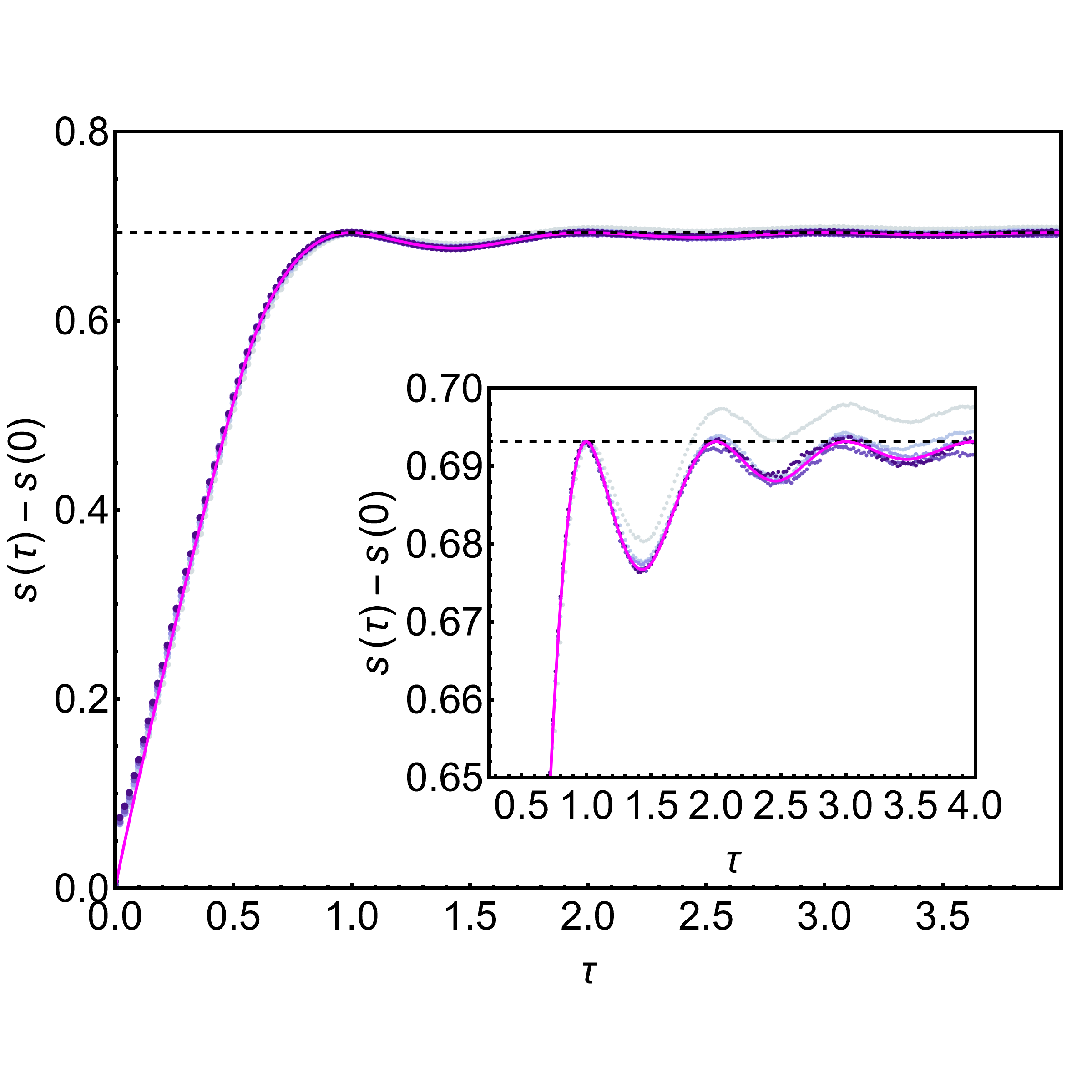}  
\includegraphics[width=7cm]{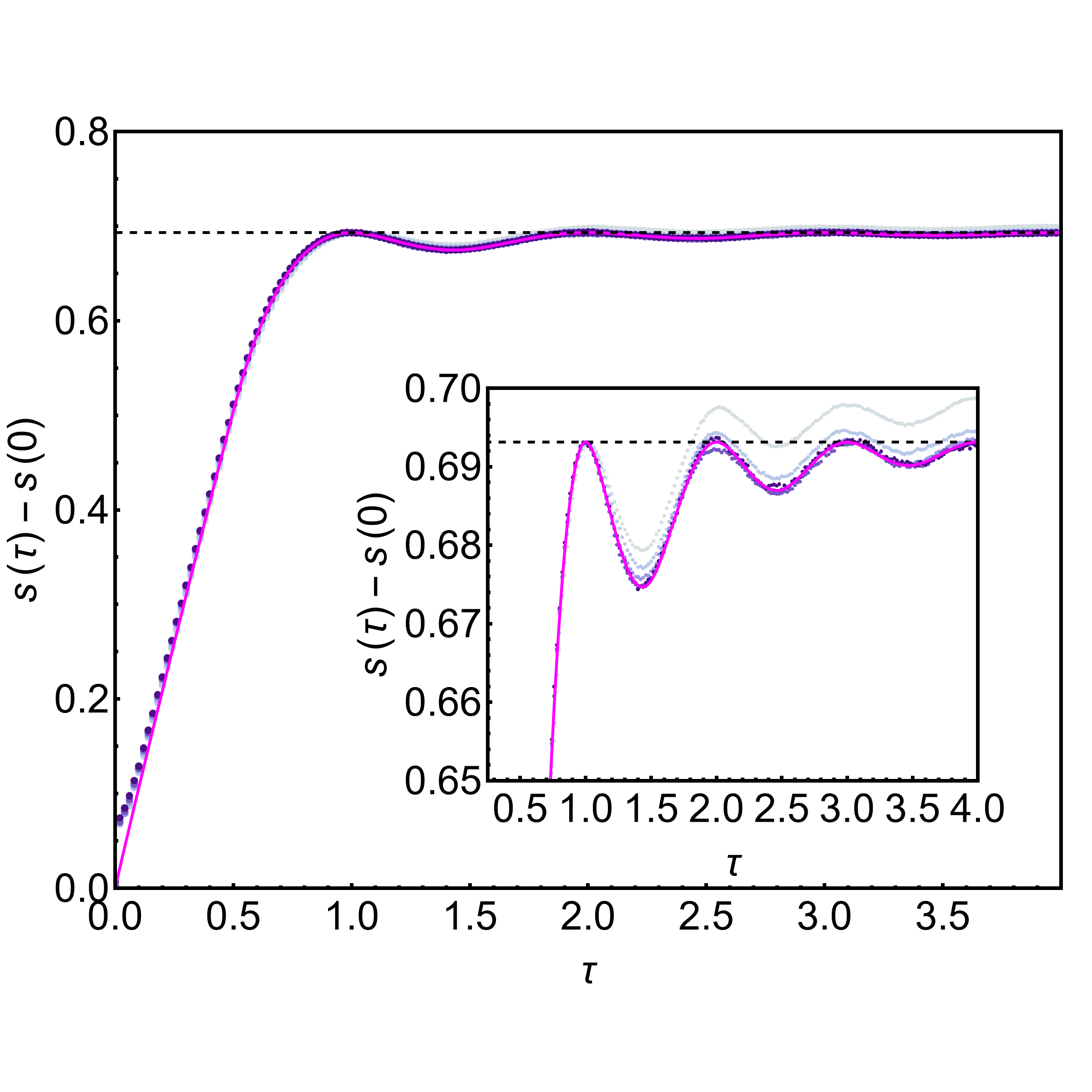}  
\captionof{figure}{Scaled entropies for hard discs. First row: $\eta_0=10^{-9}$ and $10^{-10}$. Second row: $\eta_0=10^{-11}$ and $10^{-12}$. Dots are data from simulations with $N=10^5$ discs,  $n_x=16$. Different colors are simulations with $(\Delta v_x,\eta)=(12/n_v,\eta_0 (256/n_v)^2$ with $n_v=16,32,64,128,256$ (color is darker as $n_v$ increases). Magenta lines are the solutions from eq.(\ref{en8}) with $C=\Gamma/\Delta v_x$ values obtained from the fits done in figure \ref{fignu} for each $\eta_0$.} 
\label{fit1}  
\end{center}

{\bf The insets in figure \ref{fit1} show weak apparent violations of the second law: there are time intervals where the entropy decreases.  This only occurs when the mean free path is comparable to or larger than the size of the system, so this feature does not occur if one takes the limit of a large system while keeping the mean free path fixed and finite.  This entropy decrease is due to the free motion and is present if the interparticle scattering is weak enough and $\Delta v_x$ is chosen large enough so that the entropy change due to the free motion remains larger than or comparable to that due to interparticle scattering.  The free motion component of the entropy change is not strictly subject to the second law; although it generally does produce an entropy increase for spatially inhomogeneous systems, it can produce small but extensive entropy decreases that depend in detail on the choice of $\Delta v_x$ and the time, as we have shown.  These decreases only occur when the system is dilute enough so that these features are not overwhelmed and removed by the entropy production due to the interparticle collisions.  The latter source of entropy production is, of course, fully subject to the second law.}

We have seen, and our results make explicit, that the Boltzmann entropy of a 
microstate $X$ may significantly depend on the choice of macrovariables, and in particular on the choice of cell size $\Delta v$. 
For example, $X$ may be within the equilibrium macrostate for some choice of $\Delta v$, but for a choice of a smaller $\Delta v$ it may no longer be in the equilibrium macrostate. Nonetheless, this dependence on $\Delta v$ does not arise for the equilibrium entropy---the entropy of the equilibrium macrostate: Since the dominant (equilibrium) macrostate occupies almost the entire energy surface or shell, the equilibrium entropy depends only on the volume of the energy surface or shell, up to a negligible error, regardless of the choice of macrovariables (assuming that there is a dominant macrostate such as we've assumed in this paper). Changing $\Delta v$ leads to a new equilibrium macrostate whose volume differs little from the original one. The very small differences are those microstates $X$ that are  included in the equilibrium macrostate for one value of $\Delta v$ and are not included for the other value.

\section{summary}

In this paper we have investigated the time evolution of the Boltzmann entropy of the macrostates of a dilute classical gas when the system expands freely after removing a constraint.  In numerical simulations we specifically studied a dilute two-dimensional gas of hard discs.  The macrostates are specified by dividing the one-particle phase space in to cells and counting the number of particles in each such cell.  We identify two mechanisms of entropy increase: (i) dispersal of the particles in physical space due to free motion, and (ii) dispersal of the particles in velocity space due to collisions between particles.  The first mechanism can dominate when the mean free path $\lambda$ is much larger than the length scale $\ell$ of the initial spatial inhomogeneity of the gas.  In this case the rate of the approach to the final equilibrium entropy is proportional to $(\Delta v)/\ell$, where $\Delta v$ is the size of the phase space cells in the velocity direction.  In the opposite limiting case {\bf when $(\Delta v)/\ell\ll\Gamma$, where $\Gamma$ is the interparticle scattering rate,} the rate of entropy increase becomes independent of $\Delta v$ and is given by the collision terms in the Boltzmann equation.  It was only this latter limit that Boltzmann considered in his famous H-theorem.  In the intermediate cases we show numerically, and partly analytically, that the time dependence of the entropy has a functional form set by the ratio of the rates of the two mechanisms of entropy growth.



\section{Acknowledgements}

We thank the IAS for the hospitality during part of this work. P.G.  acknowledges the support of the Spanish “Ministerio de Ciencia e Innovación” and “Agencia Estatal de Investigación”, MICIN/AEI/10.13039/501100011033,  Project Ref. PID2020-113681GB-I00.  D.A.H. was supported in part by NSF QLCI grant OMA-2120757. 

\section{Appendix: The Long-Time Behavior of the Boltzmann Entropy}

Let $F(x,v)$ be the one-particle (marginal) distribution, on $[0,L]\times \Bbb R$, of an $N$-particle system, normalized to $N$, so that 
\begin{equation}
    \int_0^L dx\, \int_{-\infty}^{\infty}dv\,  F(x,v) = N.
\end{equation}
Its {\it Gibbs entropy per particle} is given by
\begin{equation}\label{ge}
    s(F) =-\frac{1}{N} \int_0^L dx\, \int_{-\infty}^{\infty}dv\,  F(x,v)\,\log  F(x,v).
\end{equation}

A very good approximation to the Boltzmann entropy per particle  $s_B$ for a (typical) configuration arising from $F$ can be expressed in terms of $F^\Delta$, the coarse-grained one-particle distribution arising from $F$ via the coarse-graining into cells $\Delta_\alpha$ of the one-particle phase space   $[0,L]\times \Bbb R$ as described in the introduction:
\begin{equation}\label{cgf}
    F^\Delta(x,v) = F_\alpha,\quad\text{for}\ (x,v)\in \Delta_\alpha,
\end{equation}
where $F_\alpha$ is defined in \eqref{en7} (for  $v=v_x$). As stated earlier, by the law of large numbers we typically have that 
\begin{equation}\label{sbf}
    s_B \approx s(F^\Delta)
\end{equation}
in the large $N$, $\Delta\to 0$ limit.

We consider first the long-time asymptotics of $s_B(t)$ for the ideal gas in one dimension. By \eqref{sbf} we have that
\begin{equation}\label{sbft}
     s_B(t) \approx s(F_t^\Delta)
\end{equation}
where  $F_t^\Delta=(F_t)^\Delta$ is the coarse-graining of $F_t(x,v)=F(x,v,t)$, the time-evolved distribution. We thus focus here on the long-time asymptotics of $s(F_t^\Delta)$.

Note that for fixed $t$, as $\Delta\to0$,
\begin{equation}
    s(F_t^\Delta)\to s(F_t ),
\end{equation}
which is independent of $t$. However, this convergence, unlike that of the approximation \eqref{sbft}, is not uniform in $t$: For times $t$ of order $L/\Delta v$, i.e., on the time scale $\tau=t\Delta v/L$, $s(F_t^\Delta)$ does change. In fact, it is shown in \cite{dhar} that, as $\Delta\to0$,
\begin{equation}
    s(F_{\frac{\tau L}{\Delta v}}^\Delta)\to s(\bar F_\tau)
\end{equation}
with
\begin{equation}\label{gen}
    \bar F_\tau(x,v) =\frac1{\tau L}\int_{0}^{\tau L}dz \,F_0(x-z,v).
\end{equation}
(The distinction between $(F_t)^\Delta$, \z{the coarse-grained evolved distribution, and $(F^\Delta)_t$, the evolved coarse-grained distribution,} is crucial here.)

Making the changes of variables $x\to x'=x/L$ and $z\to z'=z/L$ (mapping $[0,L]\times \Bbb R$ to $[0,1]\times \Bbb R$) in \eqref{ge} and \eqref{gen}  and then writing $x$ for $x'$ and $z$ for $z'$, we obtain that
\begin{align}\label{f1}
      s(\bar F_\tau)&=  -\int_0^1 dx\, \int_{-\infty}^{\infty}dv\, \bar F_\tau^{(1)}(x,v)\,\log (\rho \bar F_\tau^{(1)}(x,v)) \\ \label{f2}
   &= -\log \rho\ - \int_0^1 dx\, \int_{-\infty}^{\infty}dv\, \bar F_\tau^{(1)}(x,v)\,\log  \bar F_\tau^{(1)}(x,v)
\end{align}
with $\rho=N/L$ and
\begin{equation}\label{bar1}
   \bar F_\tau^{(1)}(x,v) = \frac1{\tau}\int_{0}^{\tau}dz \,F_0^{(1)}(x-z,v)
\end{equation}
where
\begin{equation}
    F_0^{(1)}(x,v)= L F_0(xL,v)/N.
\end{equation}

Note that $F_0^{(1)}$, which we shall call the {\it shape} of $F_0$, is a (normalized to 1) probability distribution on 
$[0,1]\times \Bbb R$. It is the distribution of the random variable $(X/L,V)$ when $(X,V)$ is distributed according to $F_0$ on $[0,L]\times \Bbb R$. Note also that \eqref{f2} thus says that, apart from the term $-\log\rho,$ $s(\bar F_\tau)$ depends only on the shape of $F_0$ \z{---not on $L$ or $N$}.

It follows easily from \eqref{gen} that, as $\tau\to\infty$, 
\begin{equation}
    \bar F_\tau(x,v) \to F_{\text{max}}(x,v)=Ng(v)/L
\end{equation}
and hence that
\begin{equation}
    s(\bar F_\tau)\to s_{\text{max}}= s(F_{\text{max}}).
\end{equation}
Here $g$ is the global velocity distribution arising from $F_0$,
\begin{equation}\label{gv}
    g(v)= \frac1N\int_0^L dx\, F_0(x,v).
\end{equation}
For $g=g_T$, we have that $F_{\text{max}}=F_{\text{eq}}$, the one-particle equilibrium distribution at temperature $T$, and that $s_{\text{max}}=s_{\text{eq}}(\rho,T)=s(F_{\text{eq}})$, the equilibrium entropy per particle.

$s_{\text{max}}$ maximizes the entropy per particle $s(F_0)$ among all distributions $F_0$ with global velocity distribution $g$ \eqref{gv}. Thus we have for the deviation $s^{-}_\tau$ of $s(\bar F_\tau)$ from $s_{\text{max}}$,
\begin{equation}
    s(\bar F_\tau)=s_{\text{max}}-s_\tau^{-},
\end{equation}
 that $s^{-}_\tau\searrow 0$ as $\tau\to\infty$. 
 
 To explore the behavior of $s^{-}_\tau$ in more detail we write
 \begin{equation}
    F_0^{(1)}(x,v)= g(v)h(x|v).
\end{equation}
Inserting this in \eqref{bar1} we  find that 
\begin{equation}
    \bar F_\tau^{(1)}(x,v)=g(v)\bar h_\tau(x|v)
\end{equation}
with
\begin{equation}\label{barh}
   \bar  h_\tau(x|v)=\frac1\tau\int_0^\tau dz\, h(x-z|v).
\end{equation}
Then, from \eqref{f2}, we obtain that
\begin{align}
     s(\bar F_\tau)=&-\log\rho -\int_{-\infty}^\infty dv\,g(v)\log g(v)\nonumber\\ &-\int_{-\infty}^\infty dv\, g(v)\int_0^1 dx\, \bar h_\tau(x|v)\log \bar h_\tau(x|v)\\ =& \ s_{\text{max}} -\left<\int_0^1 dx\, \bar h_\tau(x|v)\log \bar h_\tau(x|v)\right>_{\!v}.\label{vavg}
\end{align}
Thus we have  that 
\begin{equation}
 s^{-}_\tau=   \left<\int_0^1 dx\, \bar h_\tau(x|v)\log \bar h_\tau(x|v)\right>_{\!v}.
\end{equation}
(Here $\left<\ \cdot\ \right>_{\!v}$ denotes the average with respect to $g(v)$.)

Note that, since (the periodic extension of) $\bar h_\tau(x|v)$ has period 1 in $x$, we have from  \eqref{barh} that $\bar h_\tau(x|v)=1$, so that  $s^{-}_\tau=0$ and $s(\bar F_\tau)=s_{\text{max}}$, for $\tau=1,2,3,\ldots.$ Note also that $s^{-}_\tau$ depends only on the shape of $F_0$, and not on $N$ or $L$.

Since the deviation of $s^{-}_\tau$ from 0 arises from that of $\bar h_\tau(x|v)$ from 1, it is convenient to write
\begin{equation}\label{hphi}
    h(x|v) = 1 + \phi(x|v).
\end{equation}
Then, from \eqref{barh}, 
\begin{equation}\label{barhphi}
   \bar h_\tau(x|v)  = 1 + \bar\phi_\tau(x|v)/\tau
\end{equation}
with
\begin{equation}\label{barphi}
    \bar\phi_\tau(x|v)=\int_0^\tau dz\,\phi(x-z|v).
\end{equation}
Note that 
\begin{equation}
  -1  \leq\bar\phi_\tau(x|v)\leq 1
\end{equation}
and, since $\int_0^1 dx\,\phi(x|v) =0$, that it is periodic in $\tau$ with period 1: For $0\leq\xi<1$ and $n=1,2,3,\ldots$
\begin{equation}
    \bar\phi_{n+\xi}(x|v)=\bar\phi_\xi(x|v).
\end{equation}

Now suppose that, as in the cases considered  in this paper, position and velocity are initially independent, so that 
\begin{equation}
    h(x|v)= h(x).
\end{equation}
Then \eqref{vavg} assumes the form
\begin{equation}
     s(\bar F_\tau)= s_{\text{max}} -\int_0^1 dx\, \bar h_\tau(x)\log \bar h_\tau(x).\label{barhx}
\end{equation}
with
\begin{equation}\label{barhtau}
    \bar  h_\tau(x)=\frac1\tau\int_0^\tau dz\, h(x-z). 
\end{equation}
Similarly, equations (39-43) of course continue to hold with $h(x)$, $\phi(x)$, $\bar h_\tau(x)$, and $\bar\phi_\tau(x)$ replacing the corresponding $v$-dependent quantities.

The choice of $h$ relevant to this paper is
\begin{equation}
    h(x)= 2\chi_{_{[0,\frac12)}}(x)
\end{equation}
with $\chi_{_{A}}$ the indicator function of the set $A$. For this $h$ we have from \eqref{hphi} that
\begin{equation}\label{diff}
    \phi(x)= \chi_{_{[0,\frac12)}}(x)-\chi_{_{[\frac12,1]}}(x),
\end{equation}
so that, from \eqref{barphi}, we find that
for $0\leq\xi\leq\frac12$,
\begin{equation}\label{barphicases}
    \bar\phi_\xi(x)\quad=\quad
    \begin{cases}
        2x-\xi, & \text{if $0\leq x\leq\xi$,} \\
        \xi, & \text{if $\xi\leq x\leq\frac12$,}\\
        -2x+1+\xi, & \text{if $\frac12\leq x\leq\frac12+\xi$,}\\
        -\xi & \text{if $\frac12+\xi\leq x\leq 1$}.
    \end{cases}
\end{equation}
Then  $\bar\phi_\xi(x)$ for $\frac12\leq\xi\leq 1$ is determined by the symmetry
\begin{equation}\label{symm}
    \bar\phi_\xi(x)=\bar\phi_{1-\xi}(1-x).
\end{equation}

In Figure \ref{b6} we display $\bar\phi_\xi$ for $0\leq\xi\leq\frac12$ and $\bar h_\tau$ for $\tau\geq 0$.
\begin{center}
\includegraphics[height=7cm]{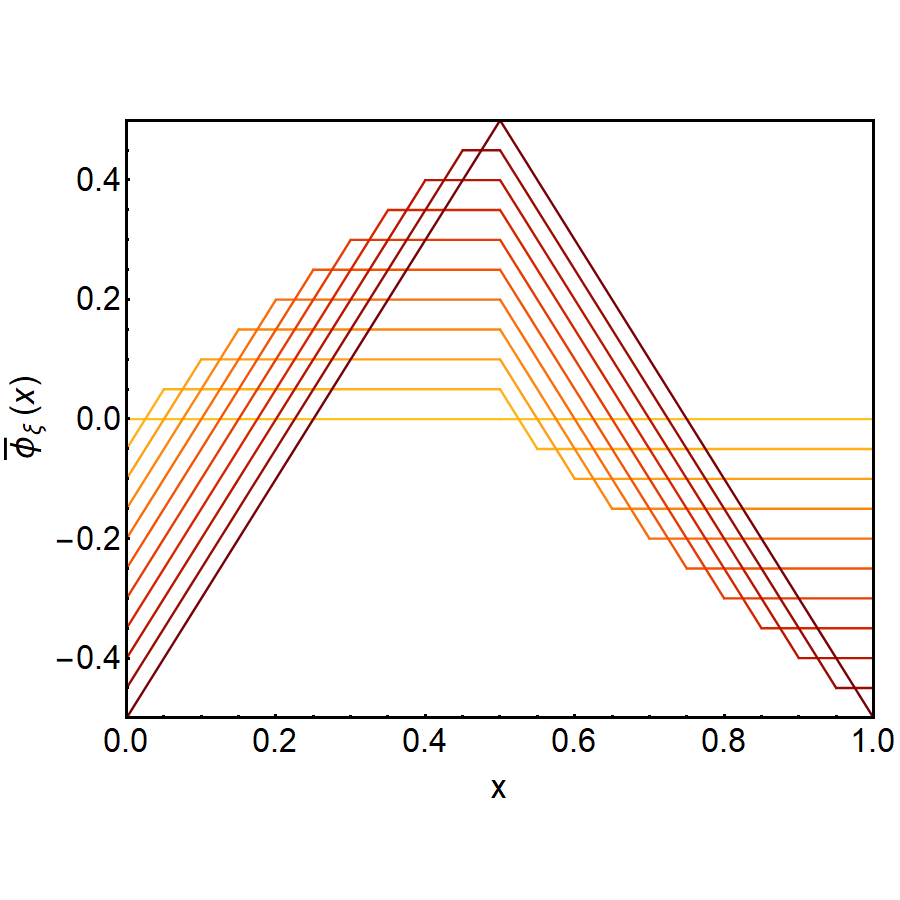}  
\includegraphics[height=7cm]{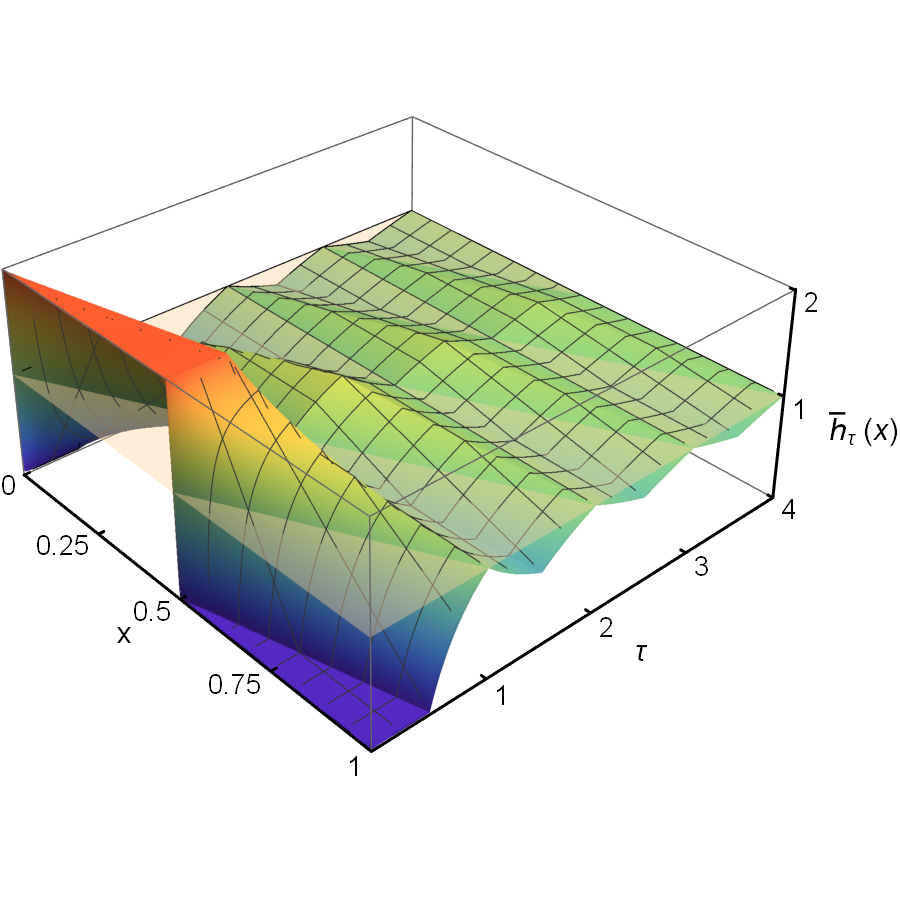}  
\captionof{figure} {Behavior of the functions $\bar\phi_\xi(x)$ and $\bar h_\tau(x)$.  Left figure are slices of $\bar\phi_\xi(x)$ for $\xi=k/20$, with $k=0,1,\ldots,10$ (color darkness increases with $k$-values). Observe that the curves for $k=11,12,\ldots 19$ are not plotted because they are just reflections about $x=1/2$ (see the symmetry property \eqref{symm}).  The horizontal plain in the right figure is  $\bar h_\tau(x)=1$.} \label{b6}  
\end{center}

In view of \eqref{symm}, we shall now write, for $n\leq\tau\leq n+1$ (with $n$ an integer), $\tau= n+\xi$ when $n\leq\tau\leq n+\frac12$ and $\tau= n+1-\xi$ when $n+\frac12\leq\tau\leq n+1$. Then  inserting \eqref{barphicases} (or its reflection via $x\mapsto1-x$) into the integral in \eqref{barhx} using \eqref{barhphi}, we obtain four  contributions, two from the intervals of length $\frac12-\xi$ on which $\bar\phi_\xi(x)$ is constant ($\pm \xi$), and two, which are the same, from the intervals on which $\bar\phi_\xi(x)$ has slope $\pm2$ and varies between $-\xi$ and $\xi$. We thus find, writing $\displaystyle{1\over \tau}=\epsilon$, that for all $\tau>0$,
\begin{align}\notag
 s(\bar F_\tau)= s_{\text{max}} - &\biggl(({\textstyle{1\over2}}-\xi)\bigl((1+\epsilon\xi) \log(1+\epsilon\xi)  + (1-\epsilon\xi) \log(1-\epsilon\xi)\bigr) \\
&+\frac1{2\epsilon}\left((1+\epsilon\xi)^2 \log(1+\epsilon\xi)-(1-\epsilon\xi)^2 \log(1-\epsilon\xi)\right)-\xi\biggr).\label{xiep}
\end{align}
Thus, for small $\epsilon$,
\begin{equation}\label{small}
    s(\bar F_\tau)=s_\text{max}-({\textstyle{1\over2}-\frac23\xi})\xi^2\epsilon^2 + O(\epsilon^4).
\end{equation}

Consider now the long-time asymptotics of $s_B(t)$ for the gas of dilute hard discs discussed in Section V. While this gas is in two dimensions, for the initial distribution of particles considered, the relevant evolution reduces to the dynamics of a gas in one dimension, with $v=v_x$. Moreover, for the approximation \eqref{eq:solrta} used there, we obtain  once again equations \eqref{xiep} and \eqref{small}, but now with $\epsilon=e^{-\tau/\bar\tau}/\tau$ instead of $\epsilon=1/\tau$. Here $\bar\tau=\Gamma^{-1}\Delta v/L$ is roughly the mean free time on the time scale $\tau=t\Delta v/L$. \z{(Note that $\epsilon=e^{-\tau_{\text{micro}}}/\tau_{\text{cg}}$ where $\tau_{\text{micro}}=\tau/\bar\tau=t/\bar t$, with $\bar t=\Gamma^{-1}$ the mean free time on the original $t$ time scale, and $\tau_{\text{cg}}=\tau=t/{\bar t_{\text{cg}}}$, with $\bar t_{\text{cg}}=L/\Delta v$ the unit for the coarse graining time scale $\tau$.)}

To see this, note that the analysis in this appendix for the case of the ideal gas covers the evolution  \eqref{eq:solrta}, provided that we replace $\phi$ in \eqref{hphi} and \eqref{diff} by $e^{-\tau/\bar\tau}\phi$, and similarly for $\bar\phi$ in \eqref{barhphi}, with \eqref{barphicases} adjusted accordingly. Note also that now $\epsilon$ (and the asymptotic entropy approximation) depends not only on $\tau$ but on $\bar\tau$ as well. Nonetheless. it continues to be the case that $s(\bar F_\tau)=s_{\text{max}}$ for $\tau=1,2,3,\ldots.$

 We show in Fig. \ref{b8} the behavior of the entropy given by eq. (\ref{xiep}) as a function of $\tau$ and the parameter $\bar\tau$.
 \begin{center}
\includegraphics[height=7cm]{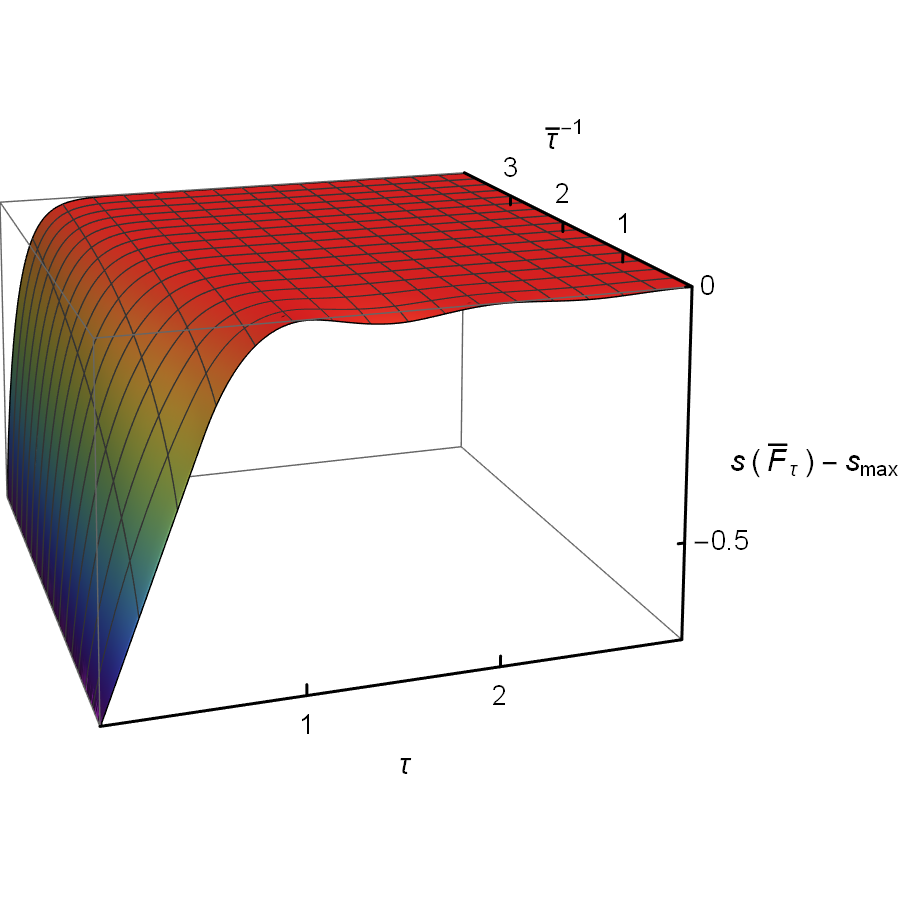}  
\captionof{figure} {Behavior of the entropy $s(\bar F_\tau)$. } \label{b8}  
\end{center}

Finally we observe  that the analysis given in this appendix and in \cite{dhar} for the one-dimensional ideal gas in $[0,L]$ applies as well to the $d$-dimensional ideal gas in $\Lambda=[0,L]^d$ (with periodic boundary conditions). Using velocity cells for the coarse-graining that are $d$-cubes aligned with the coordinate axes with edge lengths $\Delta v$, one need only replace, in the equations up to equation \eqref{barhtau}, $x\in \Bbb R$, $z\in \Bbb R$, and $v\in \Bbb R$ by $x\in \Bbb R^d$, $z\in \Bbb R^d$, and $v\in \Bbb R^d$, making the obvious adjustments for the domains of integration and the like. For example, the integration in \eqref{gen} will now be over the domain $\tau\Lambda$ (with $L$ replaced by $|\Lambda|$ in the factor in front of the integral). Similarly, the integral in \eqref{bar1} will now be over $[0,1]^d$.
Note that because $\Lambda$ and the velocity cells are similar and have the same orientation, it continues to be the case that $s(\bar F_\tau)=s_{\text max}$ for $\tau=1,2,3,\ldots$ (as it would if $\Lambda$ and the velocity cells were similar $d$-rectangles with the same orientation).

\end{document}